\pdfoutput=1
\documentclass[PAPER, atlasdraft=false, cernpreprint, UKenglish, texmf, orcidlogo]{atlasdoc}
\usepackage{atlaspackage}
\usepackage{atlasbiblatex}

\usepackage[jetetmiss=true]{atlasphysics}
\graphicspath{{logos/}{figures/}}

\usepackage{ANA-STDM-2022-02-PAPER-defs}
\usepackage{booktabs}
\usepackage{comment}
\usepackage{multicol}
\usepackage{multirow}
\usepackage{makecell}
\usepackage{siunitx}
\usepackage{cleveref}
\usepackage{hyperref}
\usepackage{catchfile}
\usepackage{placeins}
\usepackage{enumitem}

\AtlasTitle{Measurement of the cross-section for the production of a $W$ boson in
association with $b$-jets in $pp$ collisions at $\sqrt{s}=13$ TeV with the ATLAS
detector}

\AtlasAbstract{%

This paper presents a measurement of the production cross-section of a $W$ boson associated with one
or two jets, where at least one jet originates from a $b$-quark. The data are collected with the ATLAS
detector at the LHC in proton--proton collisions at $\sqrt{s}=13$ TeV, corresponding to an integrated luminosity of 140 fb$^{\text -1}$. Differential cross-sections
are presented as a function of the transverse momentum of the $b$-jet for both the electron and muon
decay modes of the $W$ boson. The results corrected for all detector effects are presented in a
fiducial region defined with basic lepton and jet kinematics, and compared with theoretical predictions.
The averaged fiducial cross-section of $W+b$-jets for the electron and muon channels is measured to
be $16.6\pm1.9$ pb, which is consistent with the next-to-leading-order QCD prediction of $16.8\pm2.3$
pb. The relative precision of this measurement is improved by approximately a factor of two compared with the previous ATLAS measurement at $\sqrt{s}=7$ TeV.

}

\AtlasRefCode{STDM-2022-02}

\PreprintIdNumber{CERN-EP-2026-146}

\AtlasDate{\today}

\AtlasJournal{J. High Energy Phys.}
\hypersetup{pdftitle={ATLAS document},pdfauthor={The ATLAS Collaboration}}

\begin{document}

\newcommand{\numround}[2]{\num[round-mode=places, round-precision=#1]{#2}}
\newcommand{\numroundsi}[2]{\num[round-mode=figures, round-precision=#1]{#2}}
\newcommand{\percentage}[2]{\SI[round-mode = places, scientific-notation = fixed, fixed-exponent = 0, output-decimal-marker={.}, round-precision = #1]{#2}}
\newcommand*\VplusBSummary[2][]{
\CatchFileDef{\tempnum}{figure/VplusBPlots/VplusBSummary/#2}{}
\unskip\unskip#1\tempnum\unskip
}

\renewcommand{\MGNLO}{\textsc{MGaMC+Py8 FxFx}\xspace}
\newcommand{\onejoneb}{\texttt{1j1b}\xspace}
\newcommand{\twojoneb}{\texttt{2j1b}\xspace}

\maketitle

\tableofcontents

\clearpage

\section{Introduction}
\label{sec:introduction}
This paper presents a measurement of the cross-section for $W$ boson production in association with one or
two jets, at least one of which originates from a $b$-quark, in proton--proton (\pp) collisions at $\sqrt{s} =$ 13 TeV.
Events in which the $W$ boson originates from a top-quark decay are excluded from the signal definition. This measurement is sensitive to heavy flavour quarks in the initial-state and provides an important test of perturbative quantum
chromodynamics (pQCD) in the presence of heavy quarks. In the Standard Model (SM), the dominant contribution to $W+b$-jets production arises from the
splitting of a gluon into a $b\bar{b}$ pair ($W+g(\rightarrow b\bar{b})$).
At next-to-leading-order (NLO), additional production mechanisms contribute, depending on the treatment of $b$-quark production~\cite{Campbell:2011udy}.

Two commonly used approaches are the four-flavour number scheme (4FNS) and the five-flavour number scheme (5FNS), which differ in the number of active quark flavours considered in the proton parton distribution functions (PDFs).
Following the notation convention of Ref.~\cite{Campbell:2011udy}, in the 4FNS the relevant processes include $q\bar{q}'\rightarrow Wb\bar{b}$($g$) and $gq\rightarrow Wb\bar{b}q'$,
while in the 5FNS additional single $b$-quark processes, such as $bq\rightarrow Wbq'$($g$) and $bg\rightarrow Wbq\bar{q}'$, also contribute significantly at
LHC energies~\cite{PhysRevD.79.034023}. A precise experimental measurement of the $W+b$-jets
production cross-section provides important input to the refinement of theoretical calculations in
pQCD, as well as the validation of Monte Carlo (MC) techniques.

The $W+b$-jets process constitutes an important background to measurements of Higgs boson production, in particular in associated production modes $WH$ and $ZH$ with $H\rightarrow b\bar{b}$ decays~\cite{HIGG-2020-20}, which are crucial for probing the coupling of the Higgs boson. The $W+b$-jets process is also an irreducible background in many searches
for physics beyond the Standard Model, and in measurements of single top-quark production~\cite{TOPQ-2018-04, Goh:2006wj}.

The first measurement of the fiducial $W+b$-jets cross-section at a hadron collider was performed by the CDF
Collaboration at the Tevatron with $\sqrt{s}$ = 1.96 TeV~\cite{CDF:2009mpr}. This measurement is in tension with the corresponding NLO theoretical prediction, which lies below the data. A subsequent measurement by the D0 Collaboration~\cite{D0:2014nkp} reported
a cross-section lower than the theoretical calculation, but in
agreement within theoretical uncertainties. The most recent measurement of $W+b$-jets production at the LHC was reported by the ATLAS
Collaboration based on 4.6 fb$^{\text -1}$  of data collected in \pp collisions at $\sqrt{s}$ = 7 TeV~\cite{STDM-2012-11}. The
ATLAS measurement exceeded the NLO theoretical prediction by 1.5 standard deviations.
The CMS Collaboration reported measurements of the related $W+bb$ process using 7 TeV and 8 TeV datasets, respectively~\cite{CMS-SMP-12-026,CMS-SMP-14-020}. The measured cross-sections were found to be consistent with the NLO theoretical predictions within the experimental and theoretical uncertainties.

In this paper, the $W+b$-jets cross-section is measured using 140 fb$^{\text -1}$ of \pp collision data collected
by the ATLAS detector during the Run 2 period~\cite{DAPR-2021-01}. The $W$ boson is reconstructed through its decay to an electron or muon,
and $b$-jet is identified by exploiting the long lifetime and the large mass of $b$-hadrons.
The $W+b$-jet production cross-section is measured in exclusive final states containing exactly one or exactly two reconstructed jets.
The one-jet category requires exactly one reconstructed jet, which must satisfy the $b$-tagging requirements (referred to as $b$-jet in the following). The two-jet category requires exactly two reconstructed jets, of which exactly one is required to be $b$-tagged.
Events with two or more reconstructed $b$-jets, as well as events with three or more reconstructed jets, are rejected in order to suppress the backgrounds from top-quark production.
%


\section{The ATLAS detector}
\label{sec:ATLASDetector}
\newcommand{\AtlasCoordFootnote}{
ATLAS uses a right-handed coordinate system with its origin at the nominal interaction point (IP)
in the centre of the detector and the \(z\)-axis along the beam pipe.
The \(x\)-axis points from the IP to the centre of the LHC ring,
and the \(y\)-axis points upwards.
Polar coordinates \((r,\phi)\) are used in the transverse plane,
\(\phi\) being the azimuthal angle around the \(z\)-axis.
The pseudorapidity is defined in terms of the polar angle \(\theta\) as \(\eta = -\ln \tan(\theta/2)\) and is equal to the rapidity
$ y = \frac{1}{2} \ln \left( \frac{E + p_z}{E - p_z} \right) $ in the relativistic limit.
Angular distance is measured in units of \(\Delta R \equiv \sqrt{(\Delta \eta)^{2} + (\Delta\phi)^{2}}\).
}

The ATLAS experiment~\cite{PERF-2007-01} at the LHC is a multipurpose particle detector
with a forward--backward symmetric cylindrical geometry and a near \(4\pi\) coverage in
solid angle.\footnote{\AtlasCoordFootnote}
It consists of an inner detector system (ID) surrounded by a thin superconducting solenoid
providing a \num{2} T axial magnetic field, electromagnetic and hadronic calorimeters,
and a muon spectrometer. The inner tracking detector covers the pseudorapidity range \(|\eta| < 2.5\).
It consists of silicon pixel, silicon microstrip, and transition radiation tracking detectors.
Lead/liquid-argon (LAr) sampling calorimeters provide electromagnetic (EM) energy measurements
with high granularity within the region \(|\eta|< 3.2\).
A steel/scintillator-tile hadronic calorimeter covers the central pseudorapidity range (\(|\eta| < 1.7\)).
The endcap and forward regions are instrumented with LAr calorimeters
for EM and hadronic energy measurements up to \(|\eta| = 4.9\).
The muon spectrometer surrounds the calorimeters and is based on
three large superconducting air-core toroidal magnets with eight coils each.
The field integral of the toroids ranges between \num{2.0} and \num{6.0}{T m}
across most of the detector.
The muon spectrometer includes a system of precision tracking chambers up to \(|\eta| = 2.7\) and fast
detectors for triggering up to \(|\eta| = 2.4\).
The luminosity is measured mainly by the LUCID--2~\cite{LUCID2} detector that is located close to the beampipe.
A two-level trigger system is used to select events~\cite{TRIG-2016-01}.
The first-level trigger is implemented in hardware and uses a subset of the detector information
to accept events at a rate close to \num{100} kHz.
This is followed by a software-based trigger that
reduces the accepted rate of complete events to \num{1.25} kHz on average
depending on the data-taking conditions.
%
%
%


\section{Data and simulated samples}
\label{sec:dataMC}
The data used in this analysis were collected with the ATLAS detector between 2015 and 2018
in \pp collisions at $\sqrt{s}$ =13 TeV. The corresponding total integrated luminosity is 140
fb$^{\text -1}$ with a relative uncertainty of 0.83\%~\cite{DAPR-2021-01}.
Data quality requirements are applied to ensure that all relevant detector subsystems were operating under stable and optimal conditions.
Data events are collected using single-electron or single-muon triggers \cite{TRIG-2018-05,TRIG-2018-01}. The lowest \pt\
threshold of the muon triggers is 26 GeV, while the lowest transverse energy (\et) threshold
used for the electron trigger is 24 GeV for the 2015 data-taking period and was raised to 26 GeV in subsequent years to
accommodate the increasing LHC instantaneous luminosity.

Monte Carlo events with full detector simulation are used to model signal and most background processes.
For inclusive $W$ boson production in association with both light- and heavy-flavour jets, two sets of fully simulated samples are used, comprising all leptonic decays $W\rightarrow \ell\nu$ ($\ell = e, \mu, \tau$). One set is produced using the
\Sherpa\ v2.2.11~\cite{Bothmann:2019yzt} ATLAS configuration~\cite{PMGR-2021-01} that performs explicit matrix
element (ME) calculations of 0--2 partons at NLO and 3--5 partons at leading
order (LO).  The PDF set used for event generation is \NNPDF 3.0 NNLO \cite{Ball:2014uwa} with $\alpha_S$ = 0.118.
The other set is generated using \textsc{MadGraph5\_aMC@NLO}~\cite{Alwall:2014hca} for $W+$0--3 partons at NLO accuracy, merged with the \FXFX\ procedure~\cite{Frederix:2012ps}, and interfaced to \PYTHIA[(v8.245)]~\cite{Sjostrand:2014zea} for parton showering and hadronisation using the A14 tune~\cite{ATL-PHYS-PUB-2014-021} (\MGNLO). For this sample, the
\NNPDF 3.1 \LUXQED NNLO \cite{Bertone:2017bme} PDF set is used, also with $\alpha_S$ = 0.118.
Both signal samples for the $W$+jets process are based on the 5FNS\@. The \Sherpa\ sample is used for the nominal measurement while the
\MGNLO\ sample is employed for cross-checks and for the evaluation of modelling uncertainties.

For $W$+jets MC samples, events are further categorised according to the flavour of the hadron associated with the jet. At particle-level, jets are classified as $b$-, $c$-, or light-jets using a cone-based matching procedure. Jets are labelled as $b$-jets if they are found within $\Delta R = 0.3$ of at least one $b$-hadron with $\pt > 5$ GeV. In cases where a $b$-hadron is matched to more than one jet, only the jet closest in $\Delta R$ is labelled as a $b$-jet. Jets not identified as $b$-jets are subsequently labelled as $c$-jets if they are within $\Delta R = 0.3$ of any $c$-hadron with $\pt > 5$ GeV. All remaining jets are classified as light-jets. Simulated events are then classified sequentially as $W+b$-jets or $W+c$-jets according to whether the leading (highest \pt) heavy-flavour jet is a $b$- or $c$-jet, respectively. Events containing only light-jets are classified as $W+$light-jets.

The $Z$+jets processes
are simulated with \Sherpa\ v2.2.11 using the same configuration as for $W$+jets. The top quark processes, including $t\bar{t}$ and single top in the $tW$, $s$- and $t$-channels, are simulated at NLO
using the \POWHEGBOX~v2~\cite{Frixione:2007nw,Nason:2004rx,Frixione:2007vw,Alioli:2010xd} generator and interfaced to \PYTHIA[(v8.230)]
for showering and hadronisation using the A14 tune. The $t\bar{t}$ sample is further reweighted to match the NNLO QCD ( + NLO EW) differential calculation~\cite{Czakon:2017wor}. The diboson ($VV$, $V$=$W,Z$) processes,
where one vector boson decays leptonically and the other decays hadronically, are simulated
using \Sherpa\ v2.2.1~\cite{Bothmann:2019yzt}, with NLO MEs for up to one parton emission and LO MEs for two or three parton emissions.

The generated samples are processed using the \GEANT-based ATLAS detector simulation~\cite{SOFT-2010-01,Agostinelli:2002hh} and the same reconstruction algorithms are applied to simulated and recorded data. To account for pile-up, multiple overlaid inelastic \pp collisions are simulated with \PYTHIA[(v8.186)] using the A3 tune~\cite{ATL-PHYS-PUB-2016-017} and the \NNPDF 2.3 LO PDF~\cite{Ball:2012cx}. The distribution of the
average number of interactions per bunch crossing in the simulations is weighted to reflect the one in
data. Simulated events are then corrected for known differences between data and simulation in reconstruction and identification efficiencies, and in energy and momentum scales and resolutions.


\section{Event reconstruction and selection}
\label{sec:obj_evt_selection}
Candidate $W+b$-jets events are required to have exactly one high-\pt\ electron or muon,
as well as missing transverse momentum consistent with the presence of a neutrino from the $W$ boson decay,
and one or two reconstructed jets, exactly one of which must be identified as a \bjet. All events must have at
least one reconstructed vertex formed by the intersection of at least two tracks with $\pt > 500$ MeV. In events with multiple vertices, the vertex with the largest $\Sigma \pt^2$  of the
associated tracks is taken to be the primary hard-scatter vertex (PV).

Electron candidates are required to satisfy $\et > 28$ GeV and $|\eta| < 2.47$, excluding the
calorimeter transition region $1.37<|\eta|<1.52$. They are reconstructed by matching topological energy clusters in the electromagnetic calorimeter with inner-detector tracks. Electron identification employs a
likelihood discriminant that combines calorimeter shower shapes, tracking
and cluster-track matching quantities. The ``Tight'' identification working point~\cite{PERF-2017-01} is used to maximise the rejection of background from misidentified objects. Additional requirements are imposed on the
electron tracks: the longitudinal impact parameter ($z_0$) is required to satisfy $|z_0\sin\theta|
< 0.5$ mm, where $\theta$ is the angle of the track with respect to the beam-line, and the transverse
impact parameter significance ($d_0/\sigma_{d_0}$) is required to satisfy $|d_0/\sigma_{d_0}|<5$.
Electrons must also fulfil the ``PLImprovedVeryTight'' isolation criterion~\cite{EGAM-2021-01} that
significantly suppresses background from photons or hadrons misidentified as electrons and non-prompt electrons from the decay of heavy-flavour hadrons.

Muon candidates are selected in the region $|\eta| < 2.5$ and are required to have $\pt > 28$ GeV.
They are identified by matching ID tracks from the PV to either full tracks or track segments
reconstructed in the muon spectrometer. Muons are required to satisfy the ``Tight'' identification working point~\cite{MUON-2018-03},
based on quality criteria applied to the inner-detector and muon-spectrometer tracks. In addition, the longitudinal impact parameter of the muon track is required to
satisfy $|z_0\sin\theta| < 0.5$ mm, and the transverse impact parameter significance is required to satisfy $|d_0/\sigma_{d_0}| < 3$. Muons must also be isolated, passing the ``Tight\_VarRad''
working point~\cite{MUON-2018-03}, which is built from tracking and calorimeter information, with a \pt-dependent
variable cone.

The presence of a neutrino in the $W$ boson decay is inferred from the missing transverse momentum. It is computed as the negative vector sum of the transverse momenta of all identified physics objects and an additional soft term. The soft term is constructed from all tracks that are associated
with the primary vertex, but not associated with any selected physics object~\cite{JETM-2020-03}. In this
analysis, the magnitude of missing transverse momentum (\met) is required to satisfy 28 GeV $<\met <$ 60 GeV. The upper
bound provides effective suppression of backgrounds from top-quark production. %

Jets are reconstructed using the anti-$k_t$ algorithm~\cite{Cacciari:2008gp} with a
radius parameter $R$ = 0.4, as implemented in the \Fastjet package~\cite{Cacciari:2011ma}. The inputs to this algorithm are particle flow (PFlow) objects~\cite{PERF-2015-09}, which
combine measurements from the ATLAS inner detector and calorimeters to improve the jet energy
resolution and increase the jet reconstruction efficiency. Jets are calibrated using a simulation-based
calibration scheme, followed by in situ corrections to account for differences between simulation and
data~\cite{PERF-2016-04}. To suppress jets originating from pile-up interactions, jets with $\pt < 60$ GeV and $|\eta| < 2.4$
are required to have a significant fraction of constituent tracks originating from the PV, as
defined by a jet vertex tagger (JVT) discriminant~\cite{ATLAS-CONF-2014-018}. Selected jets must have $\pt > 25$ GeV and
$|\eta| < 2.5$.

At reconstruction level, $b$-jets are identified using the DL1r $b$-tagging discriminant~\cite{FTAG-2019-07}.
DL1r is a deep neural network that takes as input the outputs of several lower-level taggers, including track impact parameter significance, secondary vertex likelihood, and a recurrent neural network capturing track correlations within a jet. It then combines these features to produce a single discriminant that significantly improves $b$-jet identification over previous boosted decision tree based taggers.
The 60\% tagging-efficiency working point is used, corresponding to a 60\% selection efficiency for particle-level $b$-jets, with rejection rates of 40 against $c$-jets and 2500 against light-jets.
This stringent working point is chosen to effectively suppress the large background from $W+c$/light-jets background. Selected $b$-jets must satisfy $\pt > 25$ GeV and $|\eta| < 2.5$.

To avoid the double-counting of detector signatures, an overlap-removal procedure is applied to uniquely identify the objects in an event. Jets within $\Delta R < 0.2$ of a selected lepton are removed if they are likely to originate from the lepton or a radiated photon. Subsequently, leptons within $\Delta R = 0.4$ of any remaining jet are discarded, as they are potential decay products of heavy quarks in the jet.

Signal events are selected by requiring exactly one lepton,
28 GeV $< \met <$ 60 GeV, and a transverse mass
$m_\text{T} = \sqrt{2\pt^{\ell}\met(1-\cos(\Delta\phi))} > 65$ GeV, where $\Delta\phi$ is
the azimuthal angle between the lepton and the missing transverse momentum. Events are futher required to contain one or two reconstructed jets, exactly one of which
must be $b$-tagged. The selected events are categorised into two exclusive signal regions:
\begin{itemize}
\item \onejoneb category: exactly one jet in the event, and it passes the $b$-jet requirement;
\item \twojoneb category: exactly two jets in the event, and only one of them passes the $b$-jet
requirement.
\end{itemize}

This categorisation facilitates a better control of backgrounds, thereby reducing uncertainties
in the final measurement. Due to the selection of central jets with high transverse momentum (\pt) and the usage of the tight $b$-tagging working point,
the majority of the $W+b$-jet events with two particle-level $b$-jets are reconstructed as events with one $b$-jet candidate.
As a result, although the measurement is performed using reconstructed events with only one $b$-jet,
it remains sensitive to the production of $W+b$-jets with two particle-level $b$-jets, and is therefore unfolded to the fiducial region with one or two $b$-jets, as described in Section~\ref{sec:extraction}.


\section{Background estimation}
\label{sec:background}
After event selection, the signal purity is approximately 30\% for both electron and muon channels.
Several processes can mimic the signal and contribute to the overall background.
The major background comes from $W+c$-jets and $W+$ light-jets. These events contain a genuine lepton from the $W$ boson decay and a jet that is misidentified as a $b$-jet, originating from either a $c$-quark or a light-flavour quark or gluon.
They exhibit kinematics nearly identical to the signal and can be distinguished only by examining the distribution of the $b$-tagger response.
As these processes are not modelled with sufficient accuracy by MC simulation, their contributions are estimated by using a data-driven technique referred to as the flavour-fit method \cite{STDM-2018-43}.

Additional background contributions arising from single-top,
$t\bar{t}$, and diboson processes also play an important role. These backgrounds include both a
real lepton and a real $b$-jet in their final state. Considering that these processes are
usually well modelled by simulation, their contributions are directly evaluated with fully simulated MC samples. Another type of background comes from multijet production, where one of the jets
is either misidentified as a lepton, or contains a non-prompt lepton produced in the decay of a heavy-quark.
A data-driven method named the ``matrix method''~\cite{EGAM-2019-01} is applied to evaluate the multijet contribution.
The $Z+$jets process also contributes to the selected events, primarily when
one lepton from the $Z$ boson decay is not reconstructed and a $b$-jet is present in the final state.
The $Z+$jets contribution is much smaller compared with $W+$jets or top processes, and is
quantified using MC simulations. The $W(\rightarrow\tau\nu)$+jets process is also treated as a background and estimated using MC simulations, contributing only around 1\% of the total event yields.

\subsection{The matrix method}
\label{subsec:multijets}
The matrix method is a widely used technique to estimate backgrounds containing non-prompt or misidentified (“fake”) leptons. The goal of the matrix method
is to estimate the number of events in the data sample with a prompt (``real'') lepton ($N_R$) and events
with a ``fake'' lepton ($N_F$).

Two event selection criteria are defined for leptons based on their isolation quality requirement:
\begin{itemize}
\item ``Tight'' events satisfy the nominal selection and $N_T$ corresponds to the number of events passing such requirements.
\item ``Anti-Tight'' events satisfy the nominal selection, except that the leptons are required to fail the nominal isolation criteria while passing a very loose track-based isolation requirement, in order to avoid using objects that are significantly different from the signal selection. The number of such events is denoted by $N_{!T}$.
\end{itemize}
Since the probabilities for real and fake leptons to satisfy the Tight selection differ, the number of fake-lepton events in the signal region can be determined from the observed number of data events passing each set of selection criteria.
The fake-lepton contribution is obtained by solving a system of linear equations that relates these event counts to the efficiencies for real and fake leptons to pass the Tight and Anti-Tight selections:
\begin{gather}
\label{Eq:mmatrix}
\begin{pmatrix}
$$N_{\mathrm{T}}$$ \\ $$N_{\mathrm{!T}}$$
\end{pmatrix} =
\begin{pmatrix}
$$   r & f   \\
1-r    & 1-f
$$
\end{pmatrix}
\begin{pmatrix}
$$N_{\mathrm{R}}$$ \\ $$N_{\mathrm{F}}$$
\end{pmatrix}.
\end{gather}
In the equation, $r$ and $f$ represent the efficiencies of passing the Tight selection for real and fake leptons, respectively.
MC simulated signal samples, which are corrected to data as in \cref{sec:dataMC}, are used to derive the efficiency parameter $r$. Events are required to satisfy the signal region kinematic selection and the selected lepton must be matched to a particle-level lepton.
The fake-lepton efficiency $f$, however, is measured in a region enriched in non-prompt and misidentified leptons, referred to as the fake control region, which is defined by
reversing the $\met$ and $m_\text{T}$ requirements used for the signal event selection: $\met < 28~\text{GeV}$ and
$m_\text{T} < 40$ GeV. All other requirements used for signal selection are kept for the fake control region
construction. Any residual contribution from prompt leptons (such as $W+$jets, $t\bar{t}$ and diboson)
is then subtracted using MC simulated samples. Both efficiencies $r$ and $f$ are determined as functions of the lepton $\pt$ and $\eta$, and are evaluated separately for the \onejoneb and \twojoneb signal regions.

The uncertainties associated with the matrix method are evaluated by considering systematic effects on
the ``real'' and ``fake'' lepton efficiency measurements.  The uncertainties of $r$ include: signal modelling, evaluated with the difference between the $r$-values obtained using \Sherpa\ and \MGNLO
signal samples mentioned in Section~\ref{sec:dataMC}, experimental uncertainties propagated to the
$r$ calculation; and the statistical uncertainty.

The uncertainties associated with the fake-lepton efficiency $f$ account for the choice of the fake control region and uncertainties propagated from the prompt-lepton subtraction.
The impact of the fake control region definition is evaluated using two alternative configurations.
In the first configuration, the kinematic criteria used to construct the fake control regions are modified.
The $\met$ requirement is inverted from $\met < 28~\text{GeV}$ to a ``signal-like'' region with
$\met > 28$ GeV, while the $m_\text{T}$ requirement is retained to ensure orthogonality with the signal region.
The matrix method is reapplied, and the deviation from the nominal value is assigned as an uncertainty.
Given the dependence of the ``fake'' efficiency on jet flavour, which may be different between the fake control
region and the signal region, in the second configuration the $b$-jet identification threshold in the fake control
region is relaxed from a 60\% tagging efficiency working point to a 70\% tagging efficiency working point in order to modify the jet flavour composition. The matrix method is repeated, and the resulting difference from the nominal
multijet background estimate is taken as an additional uncertainty.
To evaluate the impact of the prompt-lepton subtraction, both experimental and theoretical uncertainties
are considered for the MC samples used in the prompt-lepton subtraction and are propagated to the final
multijet estimation. The total
uncertainties in the multijet background estimation are approximately 70\% (85\%) in the electron channel, and 45\% (50\%) in the
muon channel for the \onejoneb (\twojoneb) region, with the dominant contribution arising from the variation in the $\met$ requirement.

\subsection{$W + $jets flavour fitting}\label{subsec:wjets}
The $W+c$-jets and $W +$ light-jets processes constitute the dominant backgrounds in this analysis,
exhibiting nearly identical kinematics to the signal. Their contributions, along with the signal
$W+b$-jets process, are extracted simultaneously using a $\chi^2$ fit to the $b$-tagging
discriminant (DL1r tagging score) distribution. This procedure is referred to as the flavour-fit method in the following.

With the 60\% efficiency working point used for the $b$-jet selection, events are further divided into two bins depending on the DL1r tagging score of the selected $b$-jet: the 20--60\% tagging-efficiency bin and the 0--20\% tagging-efficiency bin. To extract the signal event yields in the data, a $\chi^2$ is defined as follows:

\begin{equation}
\label{eq:chi2}
\chi^2 = \sum \limits_{i} \limits^{N_{\text{bins}}} \frac{(N_i^{\text{data}} -
p_0 \cdot N_i^{\text{MC,W+b}} - p_1 \cdot N_i^{\text{MC,W+j}} - N_i^{\text{MC,other}})^2}{\sigma^2_{\text{data}} + \sigma^2_{\text{MC}}}.
\end{equation}

Here $N_i^{\text{data}}$ is the number of data events in the $i$th bin and $N_i^{\text{MC,W+b}}$ is the number of signal events in the $i$th bin from MC simulation. The $W+c$-jets and $W+$ light-jets events are grouped together, denoted by $N_i^{\text{MC,W+j}}$, given that their contributions in the 0--20\% tagging bin are very small, and all other background events are similarly combined to $N_i^{\text{MC,other}}$. The scale factors, $p_0$ and $p_1$, are free parameters determined from the fit to data to extract the observed $W+b$-jets signal and combined $W+c$-jets and $W+$ light-jets background (denoted by $W+$ other-jets in the following text). The uncertainties of data or MC ($\sigma_{\text{data}}$ and $\sigma_{\text{MC}}$) do not affect the solution of this linear system. The flavour-fit procedure is performed inclusively for the total cross-section measurement, and independently in each $b$-jet \pt\ bin for the differential cross-section measurement.

Jet-flavour-dependent DL1r calibrations are applied to correct for differences in tagging efficiencies between data and simulation before the flavour fit. For the $W+b$-jets signal, corrections are derived separately in the two tagging bins following standard ATLAS procedures~\cite{FTAG-2018-01}.
For the $W+c$-jets and $W+$ light-jets backgrounds, the 0--60\% correction factors from ATLAS standard recommendations~\cite{FTAG-2019-02, FTAG-2020-08} are applied in both regions. To account for the potential differences in the correction factors in the 0--20\% region, an additional 100\% uncertainty is assigned.

The inclusive fit results for signal and $W + $ other-jets are summarised in \cref{tab:Wb_SF}.
The statistical uncertainties on the fitted parameters $p_0$ and $p_1$ are estimated with the
bootstrap method \cite{ATL-PHYS-PUB-2021-011}. One thousand replicas are generated for both data and simulated spectra based on
their statistical uncertainties, and the fit is conducted for each replica. The standard
deviation of the fitted parameters of all replicas is quoted as the statistical uncertainty.
Systematic uncertainties on the fitted parameters are evaluated using an offset method. This method
reruns the fit while varying only one systematic source at a time, and iterates through all
systematic sources. For each variation, the fitted parameters are extracted, and the deviation
from the nominal value is assigned as the systematic uncertainty. An additional systematic
uncertainty is evaluated to account for differences between \Sherpa\ and \MGNLO in modelling
$W +$jets production. This uncertainty is incorporated by substituting the DL1r score shape
of $W + $jets from one generator with the other during the flavour fitting procedure. For the
$W + b$-jets signal process, in both \onejoneb and \twojoneb signal regions, and for both electron and muon final states, the dominant
systematic uncertainties on the fitted parameters arise from $W + $jets theoretical
uncertainties and $b$-tagging calibration.  Similarly, for the $W + $  other-jets process,
the dominant systematic uncertainties on the fitted parameters arise from $b$-tagging
calibration.

\FloatBarrier
\begin{table}[htbp]
\centering
\caption{Fitted scale factors for $W + b$-jets and $W +$ other-jets in the \texttt{1j1b} and \texttt{2j1b} regions for both electron and muon final states. The quoted uncertainties correspond to the statistical (first) and systematic (second) components.}
\small
\begin{tabular}{l| c| c| c| c}
\hline
Type                     & $e$ \texttt{1j1b}        & $e$ \texttt{2j1b}        & $\mu$ \texttt{1j1b}      & $\mu$ \texttt{2j1b}      \\
\hline
$W + b$-jets ($p_0$)     & $1.02 \pm 0.01 \pm 0.16$ & $0.99 \pm 0.01 \pm 0.21$ & $0.99 \pm 0.01 \pm 0.15$ & $0.97 \pm 0.01 \pm 0.19$ \\
\hline
$W +$ other-jets ($p_1$) & $0.93 \pm 0.01 \pm 0.17$ & $1.04 \pm 0.01 \pm 0.26$ & $0.93 \pm 0.01 \pm 0.18$ & $1.02 \pm 0.01 \pm 0.26$ \\
\hline
\end{tabular}
\label{tab:Wb_SF}
\end{table}

\begin{figure}[htbp]
\centering
\subfloat[]{
\includegraphics[width=0.48\textwidth]{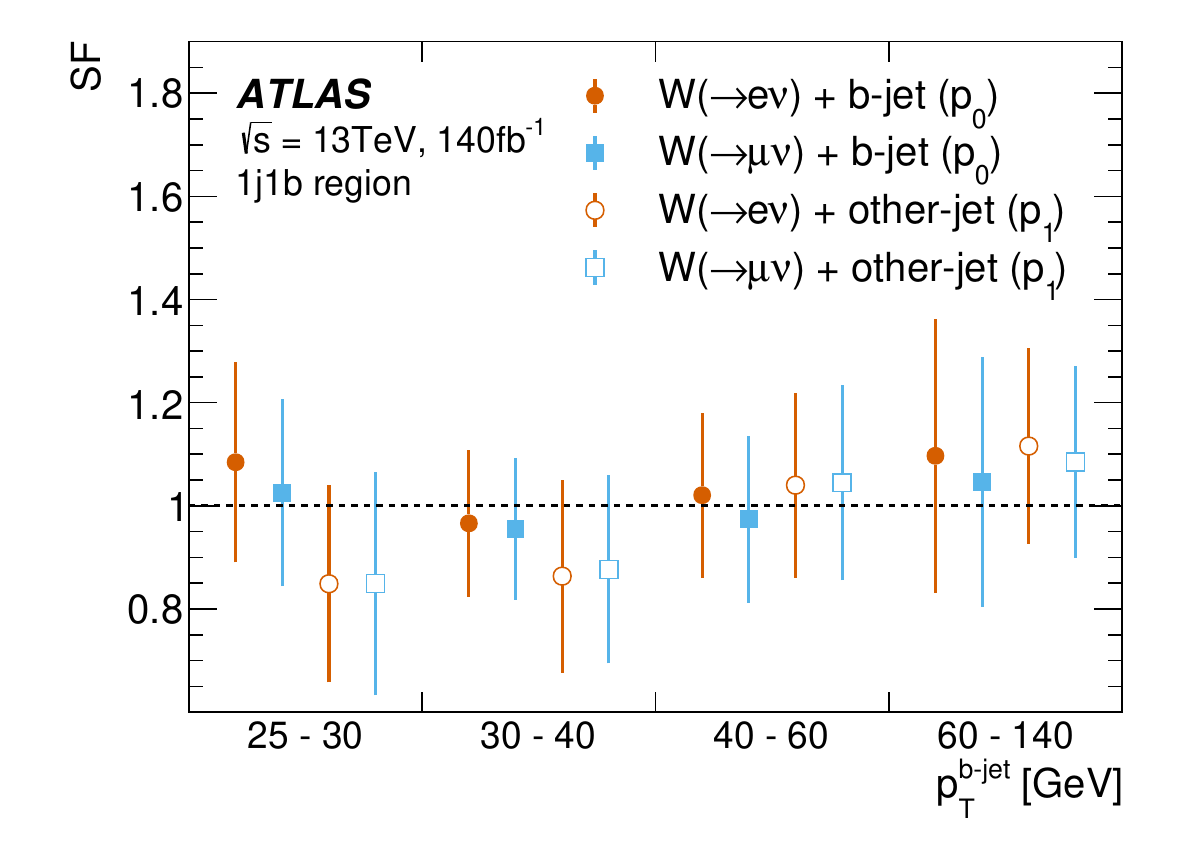}
\label{fig:WPlusB_1j1t_SherpaVjet_SF_BJetPt}
}
\subfloat[]{
\includegraphics[width=0.48\textwidth]{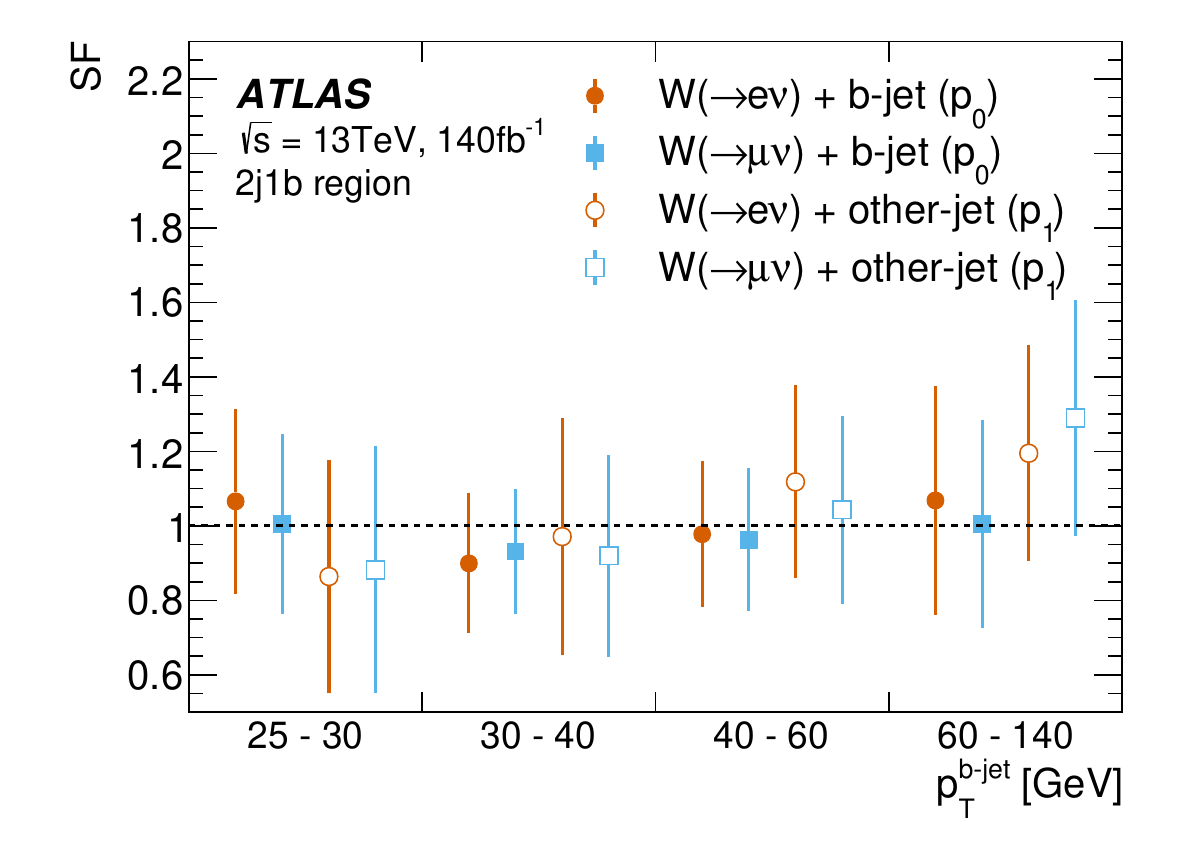}
\label{fig:WPlusB_2j1t_SherpaVjet_SF_BJetPt}
}
\caption{
Fitted scale factors (SF) for $W + b$-jets ($p_0$) and $W +$ other-jets ($p_1$) in bins of $b$-jet \pt\ in the \protect\subref{fig:WPlusB_1j1t_SherpaVjet_SF_BJetPt} \onejoneb and \protect\subref{fig:WPlusB_2j1t_SherpaVjet_SF_BJetPt} \twojoneb regions. The error bars represent the total uncertainty.
}
\label{fig:WPlusB_SF_BJetPt}
\end{figure}

For differential measurement, independent fits are conducted in four bins of $b$-tagged jet $\pt$: 25--30 GeV, 30--40 GeV, 40--60 GeV and 60--140 GeV. The fitting strategy and uncertainty evaluation are identical to those used for the inclusive measurement. The fitted scale factors are shown in \cref{fig:WPlusB_SF_BJetPt}.

\subsection{Estimated Background Contribution}
\begin{figure}[htbp]
\centering
\subfloat[]{
\includegraphics[width=0.45\textwidth]{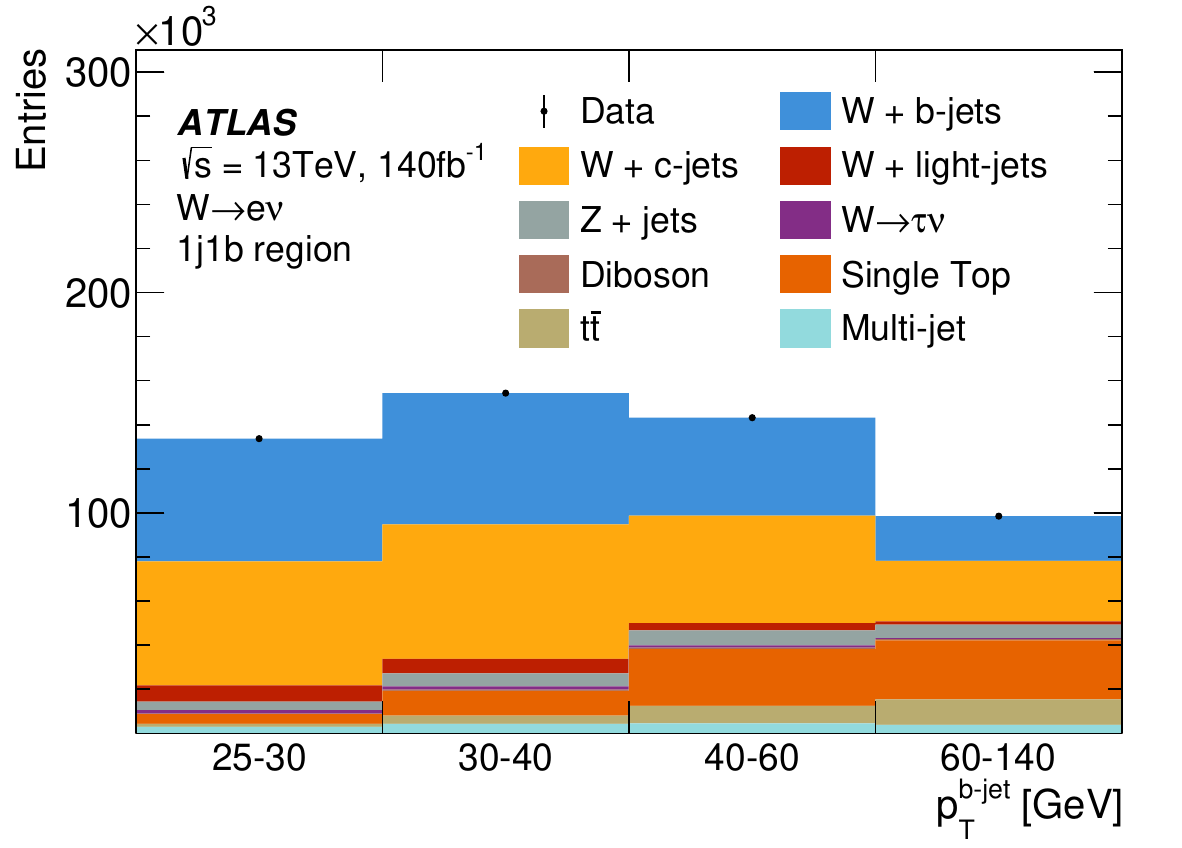}
\label{fig:Wenu_BJetPt_1j1t_Sherpa}
}
\subfloat[]{
\includegraphics[width=0.45\textwidth]{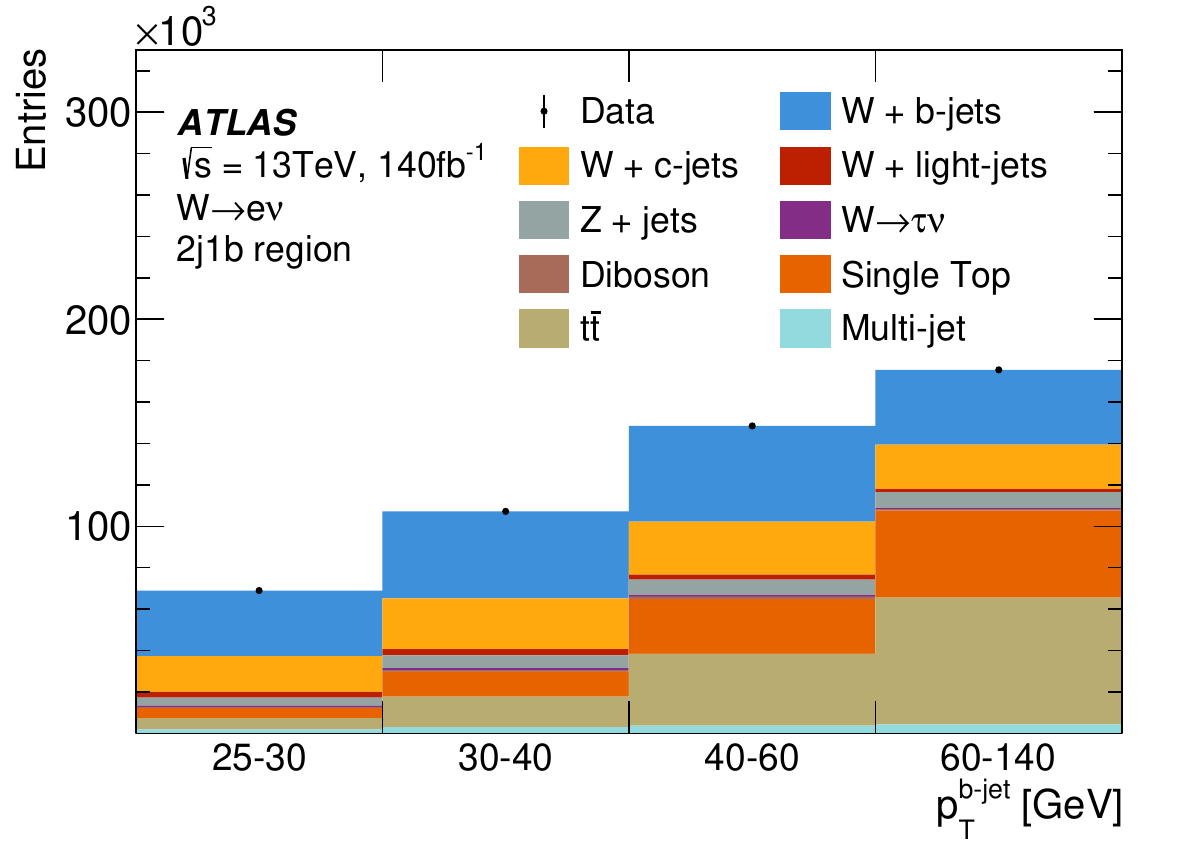}
\label{fig:Wenu_BJetPt_2j1t_Sherpa}
}
\\
\subfloat[]{
\includegraphics[width=0.45\textwidth]{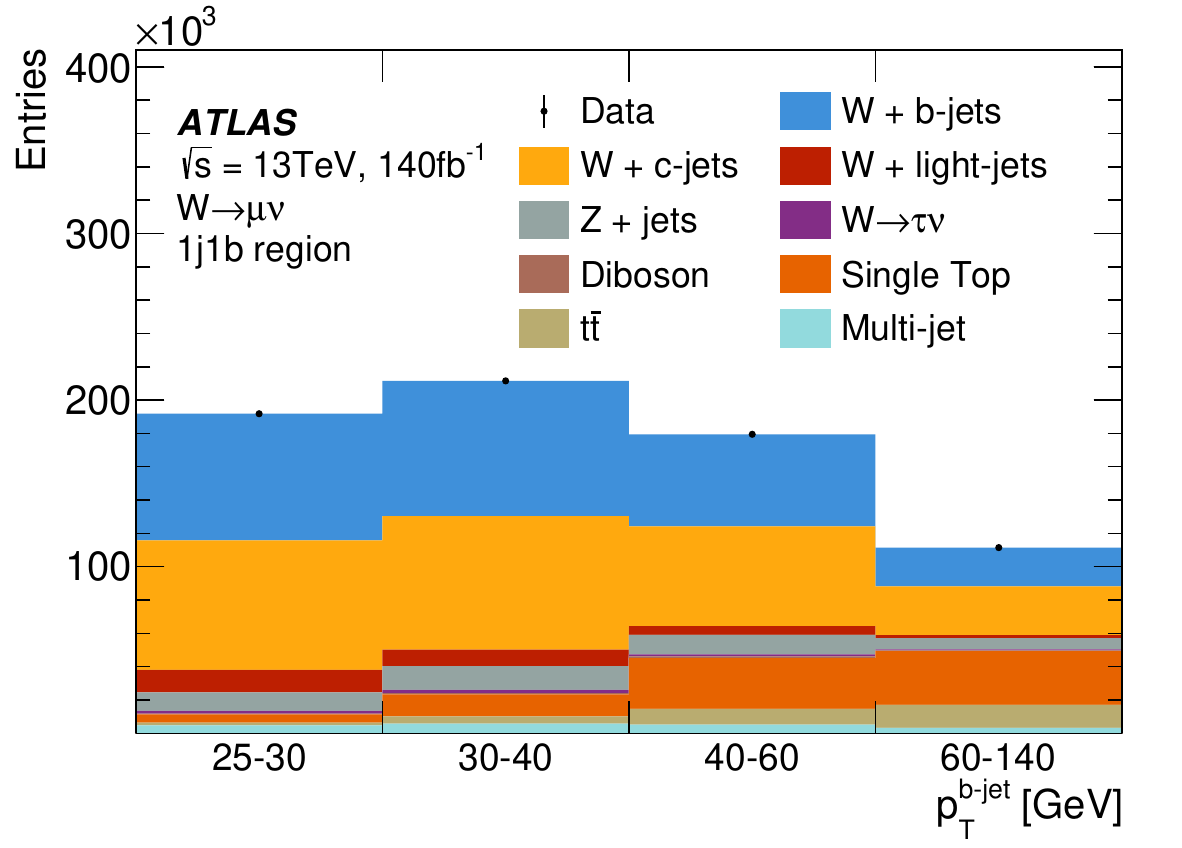}
\label{fig:Wmunu_BJetPt_1j1t_Sherpa}
}
\subfloat[]{
\includegraphics[width=0.45\textwidth]{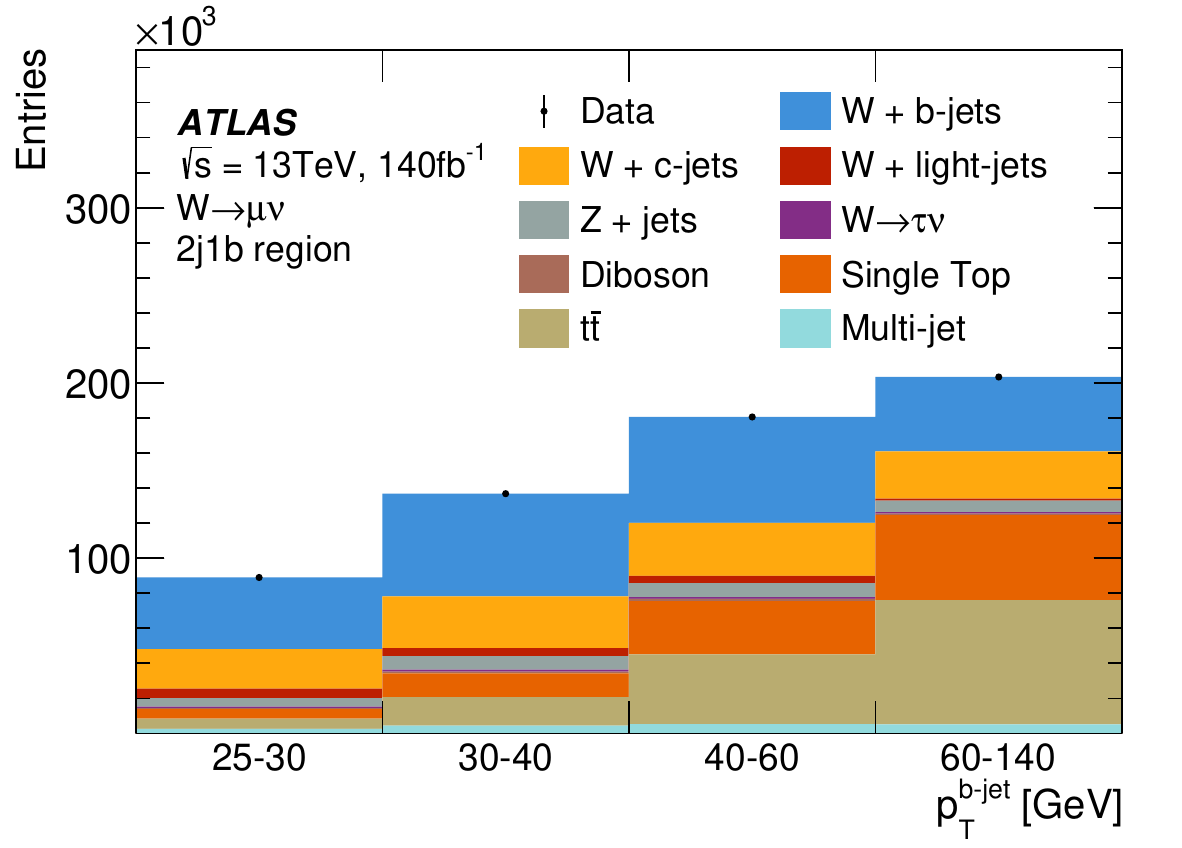}
\label{fig:Wmunu_BJetPt_2j1t_Sherpa}
}
\caption{
$b$-jet \pt\ distribution after the flavour fit for the \protect\subref{fig:Wenu_BJetPt_1j1t_Sherpa} electron \onejoneb, \protect\subref{fig:Wenu_BJetPt_2j1t_Sherpa} electron \twojoneb, \protect\subref{fig:Wmunu_BJetPt_1j1t_Sherpa} muon \onejoneb and \protect\subref{fig:Wmunu_BJetPt_2j1t_Sherpa} muon \twojoneb regions.
}
\label{fig:BJetPt_Sherpa}
\end{figure}

The expected numbers of events for all processes after the previous background estimation are summarised in \cref{tab:selevts}. In the table, $W+b$, $W+c$ and $W+$ light-jet contributions are separated according to the particle-level flavour of the $b$-tagged jet as explained in \cref{sec:dataMC}, and their event yields are corrected by the flavour fitting. The corrected $b$-jet \pt distributions are shown in \cref{fig:BJetPt_Sherpa}, demonstrating the contribution from each process.

\begin{table}[htbp]
\centering
\caption{Observed and expected number of events in each signal region after the flavour fit. Only MC statistical uncertainties are shown for the $V$+jets, diboson and top-quark processes, as well as the total predictions. For the multijet process, the total uncertainty is displayed instead.}
\begin{tabular}{l
| r@{\(\,\pm\,\)}r
| r@{\(\,\pm\,\)}r
| r@{\(\,\pm\,\)}r
| r@{\(\,\pm\,\)}r}
Type                   & \multicolumn{2}{c|}{$e$ \texttt{1j1b}} & \multicolumn{2}{c|}{$e$ \texttt{2j1b}} & \multicolumn{2}{c|}{$\mu$ \texttt{1j1b}} & \multicolumn{2}{c}{$\mu$ \texttt{2j1b}}                                  \\
\hline
$W + b$-jet            & 180100                                 & 800                                    & 159800                                   & 600                                     & 236400 & 800  & 206900 & 600   \\
$W + c$-jet            & 195000                                 & 1000                                   & 90600                                    & 600                                     & 248000 & 1000 & 110800 & 700   \\
$W +$ light-jet        & 19700                                  & 700                                    & 10500                                    & 400                                     & 32000  & 2000 & 17000  & 1000  \\
\hline
$W\rightarrow \tau\nu$ & 4500                                   & 300                                    & 3200                                     & 200                                     & 5500   & 300  & 4000   & 200   \\
$Z+$jet                & 23200                                  & 200                                    & 26100                                    & 200                                     & 44100  & 700  & 27800  & 400   \\
Diboson                & 2100                                   & 20                                     & 3150                                     & 30                                      & 2610   & 20   & 3800   & 30    \\
Single top             & 69000                                  & 90                                     & 88400                                    & 100                                     & 81900  & 100  & 101000 & 100   \\
$t\bar{t}$             & 25340                                  & 70                                     & 121000                                   & 200                                     & 29900  & 80   & 138200 & 200   \\
\hline
Multijet               & 17000                                  & 12000                                  & 15000                                    & 13000                                   & 20000  & 9000 & 20000  & 10000 \\
\hline
Total prediction       & 536000                                 & 1000                                   & 517000                                   & 1000                                    & 700000 & 3000 & 627000 & 2000  \\
Data                   & \multicolumn{2}{c|}{535566}            & \multicolumn{2}{c|}{517204}            & \multicolumn{2}{c|}{699516}              & \multicolumn{2}{c}{627216}                                               \\
\hline
\end{tabular}
\label{tab:selevts}
\end{table}

\FloatBarrier


\section{Cross-section extraction}
\label{sec:extraction}
The $W+b$-jets yields obtained from the flavour fits are converted into a fiducial cross-section for $W+b$-jets production times branching ratio for a single leptonic decay channel
($W\rightarrow\ell\nu$) by correcting for all detector effects using MC simulation. The particle-level fiducial region is defined in \cref{tab:fiducialdef}. At particle-level, objects are selected using requirements closely matching the reconstruction-level selection described in \cref{sec:obj_evt_selection} in order to minimize the dependence of the measurement on theoretical predictions. In this definition, the lepton kinematic variables are computed using final-state leptons (electrons or muons) originating from the $W$ boson decay, excluding electrons or muons from $\tau$-lepton decays. Photons radiated within a cone of $\Delta R = 0.1$ around the direction of a final-state lepton are added to the lepton four-momentum. The resulting object is referred to as a `dressed' lepton. Particle-level jets are reconstructed from generated events without detector simulation. They are clustered using the anti-$k_t$ algorithm~\cite{Cacciari:2008gp} with a radius parameter $R$ = 0.4. All detector-stable particles with a lifetime $\tau$ satisfying $c\tau > 10 \text{mm}$ are used, excluding those expected to deposit little or no energy in the calorimeters (i.e., muons and neutrinos). Particle-level jets are classified as $b$-jets using procedure described in \cref{sec:dataMC}. Particle-level selections on neutrino \pt\ and $m_\text{T}$ are looser than the reconstruction-level variants due to the large experimental resolution in these quantities.

For the inclusive measurement, two truth (particle-level) jet regions are defined for the unfolding procedure: the one-jet region (requiring one truth $b$-jet) and the two-jet region (requiring at least one truth $b$-jet). For the differential measurement, each truth region is further subdivided into four truth $b$-jet $p_{\text{T}}$ bins: 25--30 GeV, 30--40 GeV, 40--60 GeV, and 60--140 GeV. If two $b$-jets are present in an event, the leading $b$-jet (higher \pt) is used for the differential measurement.
The differential measurement is not extended to higher $b$-jet \pt\ because of the background contamination from top processes.

\begin{table}[htbp]
\centering
\caption{Definition of the particle-level fiducial region of the measurement.}
\begin{tabular}{l|c|c}
\hline
Requirement                & \multicolumn{2}{c}{Cut}                                          \\
\hline
Lepton transverse momentum & \multicolumn{2}{c}{$\pt > 28~\text{GeV}$}                        \\
Lepton pseudo-rapidity     & \multicolumn{2}{c}{$|\eta| < 2.5$}                               \\
Neutrino                   & \multicolumn{2}{c}{$\pt > 20~\text{GeV}$}                        \\
$m_\text{T}$               & \multicolumn{2}{c}{$m_\text{T} > 40~\text{GeV}$}                 \\
\hline
Jet transverse momentum    & \multicolumn{2}{c}{$\pt > 25~\text{GeV}$}                        \\
Jet rapidity               & \multicolumn{2}{c}{$|y| < 2.5$}                                  \\
Jet multiplicity           & $n=1$                                               & $n=2$      \\
$b$-jet multiplicity       & $n_b=1$                                             & $n_b\ge 1$ \\
\hline
Jet isolation              & \multicolumn{2}{c}{$\Delta R(\ell,\text{jet})>0.4$}              \\
\hline
\end{tabular}
\label{tab:fiducialdef}
\end{table}

The unfolding procedure can be expressed as the following formula:
\begin{equation}
\sigma_i=\frac{\sum_{j} A_{i,j}^{-1} \cdot f_j \cdot (d_j-b_j)}{L\cdot e_i},
\end{equation}
where the factors are defined as:

\begin{itemize}
\item $\sigma_i$: unfolded cross-section in \textit{i}-th bin in $b$-jet \pt\ or jet multiplicity.
\item $d_j - b_j$: background-subtracted data event number in \textit{j}-th bin.
\item $L$: integrated luminosity.
\item Response matrix $A$: bin-to-bin migrations between the reconstructed and the truth distribution.
\item Efficiency correction factors $e_i = N^{RT_i} / N^{T_i}$: the ratio of truth-reco-matched events ($N^{RT_i}$) to truth events ($N^{T_i}$) in the \textit{i}-th bin.
\item Fiducial correction factors $f_j = N^{R_j T} / N^{R_j}$: the ratio of truth-reco-matched events ($N^{R_j T}$) to reco events ($N^{R_j}$) in the \textit{j}-th bin.
\end{itemize}

The Monte Carlo sample simulated with \Sherpa\ is used to derive the efficiency correction factors $e_i$, the fiducial correction factors $f_j$, and the response matrix $A_{i,j}$. Alternative results are also obtained with \MGNLO\ and serve as an estimation of modelling uncertainty.
For the inclusive measurement, the matrix inversion is conducted straightforwardly, since there are only two bins in the jet multiplicity and the migration effect is small. For the differential measurement, there are eight bins in total (two jet multiplicity bins $\times$ four $p^{\text{b-jet}}_{\text{T}}$ bins), and the response matrix is inverted using an iterative Bayesian technique~\cite{DAgostini:1994fjx} to account for the low purity, mostly due to the resolution in the $p^{\text{b-jet}}_{\text{T}}$ measurement. In the Bayesian unfolding procedure, the MC prediction is used as the initial prior, and five successive iterations are performed to suppress the bias from the initial distribution. The modelling uncertainties and the potential unfolding bias will be discussed in \cref{sec:uncertainties}.


\section{Uncertainties}
\label{sec:uncertainties}
Uncertainties in the $W+b$-jets cross-section measurements arise from both statistical effects associated with the finite size of the analysed data and systematic effects related to detector-level selection, background determination, theoretical predictions and the unfolding method. Uncertainties are treated as correlated, when appropriate, between lepton channels, signal and background processes and across observables. Some uncertainty sources can affect multiple components of this analysis; these are evaluated by repeating the full analysis under each systematic variation, and taking the difference between the nominal result and the varied result as the uncertainty. The total systematic uncertainties on the final unfolded cross-sections are obtained by summing in quadrature all individual contributions.

\subsection{Detector-level systematic uncertainties}
Detector-related systematic uncertainties are defined as uncertainties arising from the selection or calibration of the physics objects entering the measurement and from the luminosity measurement. They are derived by propagating each systematic shift from each systematic source through both the flavour fit and the unfolding inputs (response matrices, efficiency correction factors, and fiducial correction factors) into the unfolded data.

Systematic uncertainties on the leptons are related to the trigger, reconstruction, identification and isolation criteria adopted in the selection of electrons and muons~\cite{EGAM-2021-01, MUON-2018-03}. Variations in the electron energy scale and resolution are taken into account, as are those related to the muon momentum scale, the inner-detector and muon spectrometer resolution, and the sagitta-bias corrections. These uncertainties are specific for each leptonic final state and are therefore treated as uncorrelated between the two lepton channels.

The uncertainties associated with the reconstructed jets take into account corrections to the energy scale and resolution~\cite{PERF-2016-04}. They are evaluated by scaling and smearing the jet four-momentum in the simulation according to the calibration uncertainties. The uncertainty in the JVT efficiency is also included. The systematic uncertainty on \met\ accounts for the energy scale and resolution of the soft hadronic activity in the event~\cite{JETM-2020-03}.

Uncertainties related to flavour tagging are derived from the calibration of the DL1r tagger using data control samples enriched in $b$-~\cite{FTAG-2018-01}, $c$-~\cite{FTAG-2020-08}, or light-jets~\cite{FTAG-2019-02}, covering jet transverse momenta up to several hundred GeV.
Discrepancies in flavour-tagging efficiencies between data and simulation are corrected in simulation as a function of jet flavour, $b$-tagging discriminant thresholds, and jet $\pt$.
The uncertainties in the flavour-tagging calibration are extrapolated to high \pT using MC simulation \cite{ATL-PHYS-PUB-2021-003}.

The uncertainty in the combined 2015--2018 integrated luminosity is 0.83\%~\cite{DAPR-2021-01}, obtained using the LUCID-2 detector for the primary luminosity measurements~\cite{LUCID2}.

\subsection{Background and theoretical systematic uncertainties}
In addition to detector effects, uncertainties from imperfect MC modelling are included for both signal and background processes. For top quark processes, the modelling uncertainties considered in this analysis include variations in NLO matching, parton shower, initial- and final-state radiation, and resummation damping. The theoretical uncertainties due to missing higher-order corrections are probed by variations of the QCD scales, the PDF set, and $\alpha_S$. These effects are evaluated using alternative MC samples or modified parameter values following standard ATLAS prescriptions~\cite{TOPQ-2018-26}.

Theoretical uncertainties due to missing higher-order corrections in $V$+jets production arise from two main sources. The scale uncertainties are determined by varying independently the QCD renormalisation ($\mu_R$) and factorisation ($\mu_F$) scales by a factor of 0.5 or 2 with an additional constraint of $0.5 \le \mu_R/\mu_F \le 2$. The envelope of the varied predictions is taken as the error estimate, commonly referred to as the ``7-point uncertainty''. Effects due to PDF uncertainties are also considered and follow the recipes provided by the \NNPDF\ group~\cite{Ball:2014uwa}.

For signal $W+b$-jet processes, an additional modelling uncertainty is considered by comparing \Sherpa\ and \MGNLO\ generators. This variation affects both the flavour fitting and the unfolding procedure. The impact of this modelling uncertainty is evaluated by repeating the full analysis using the \MGNLO\ samples and taking the difference with respect to the nominal \Sherpa-based result.

The estimations of uncertainties associated with the multijet background and with the $W+c$-jets and $W+$ light-jets backgrounds are described in \cref{subsec:multijets} and \cref{subsec:wjets}, respectively.

\subsection{Unfolding systematic uncertainties}
Uncertainties in the unfolding procedure arise from the statistical uncertainty of \Sherpa\ signal MC samples, from the modelling uncertainties of the \Sherpa\ signal samples, and from the intrinsic bias of the unfolding method associated with the prior used in the Bayesian unfolding procedure.

The impact of the finite simulated sample size is evaluated using 1000 pseudo-experiments, in which the unfolding inputs are fluctuated independently according to Gaussian distributions and the standard deviation of the results is adopted as the uncertainty estimate.

Signal theoretical uncertainties arising from missing higher-order corrections and other modelling
uncertainties can affect response matrices, efficiency correction factors and fiducial correction factors. These uncertainties are evaluated using QCD scale and PDF variations in the \Sherpa\ $W+$jet sample, as well as by comparing the unfolded results with those obtained using \MGNLO.

The dependence of the Bayesian unfolding regularisation procedure on the choice of the initial prior is determined by reweighting the generator-level distribution of $p^{\text{b-jet}}_{\text{T}}$ in the \Sherpa\ samples to provide a better description of the data at detector level. The reweighted samples are treated as pseudo-data and unfolded using the nominal response matrix, and the observed deviation from the reweighted generator-level distribution is quoted as the unfolding bias introduced by the regularisation.

The statistical uncertainty of the data is propagated through the unfolding by using 1000 pseudo-experiments. The flavour fit and the unfolding are repeated for each data replica.


\section{Results}
\label{sec:results}
\begin{figure}[htbp]
\centering
\subfloat[]{
\includegraphics[width=0.48\textwidth]{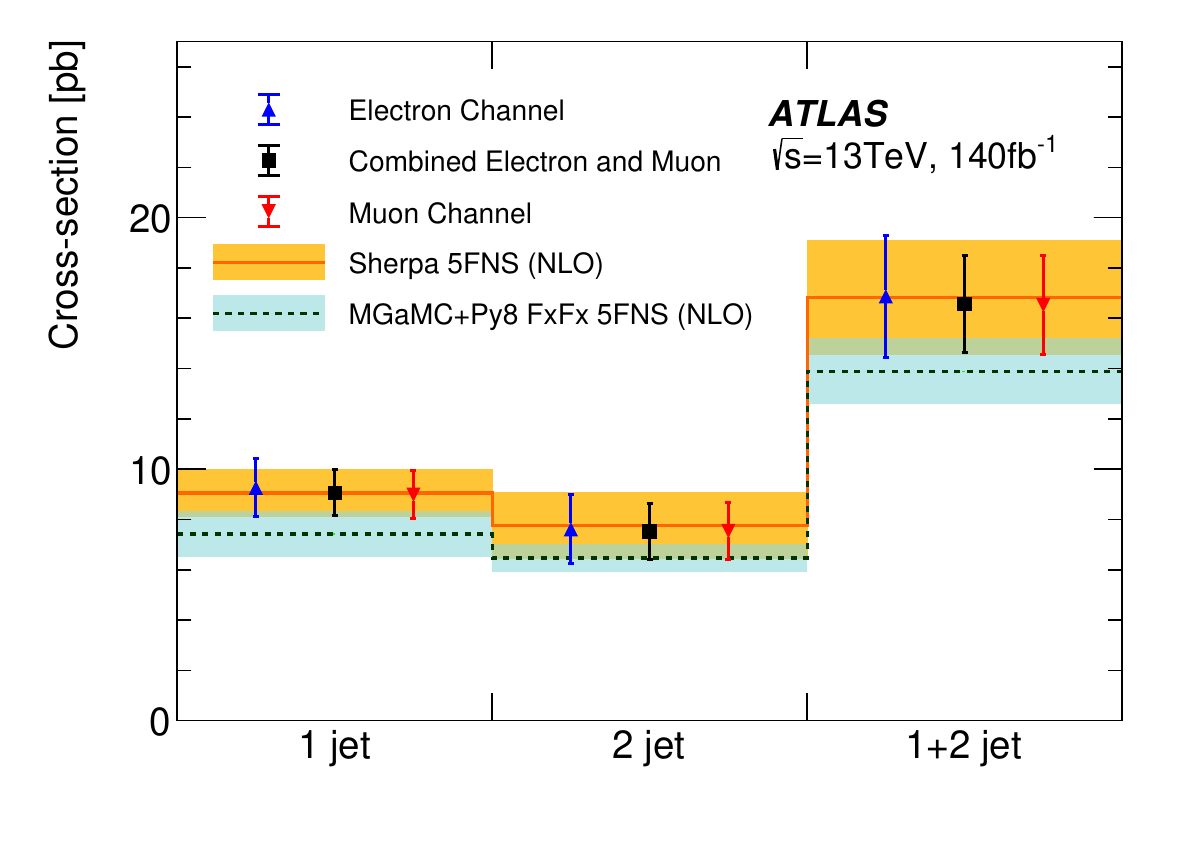}
\label{fig:TotalXsec}
}
\centering
\subfloat[]{
\includegraphics[width=0.48\textwidth]{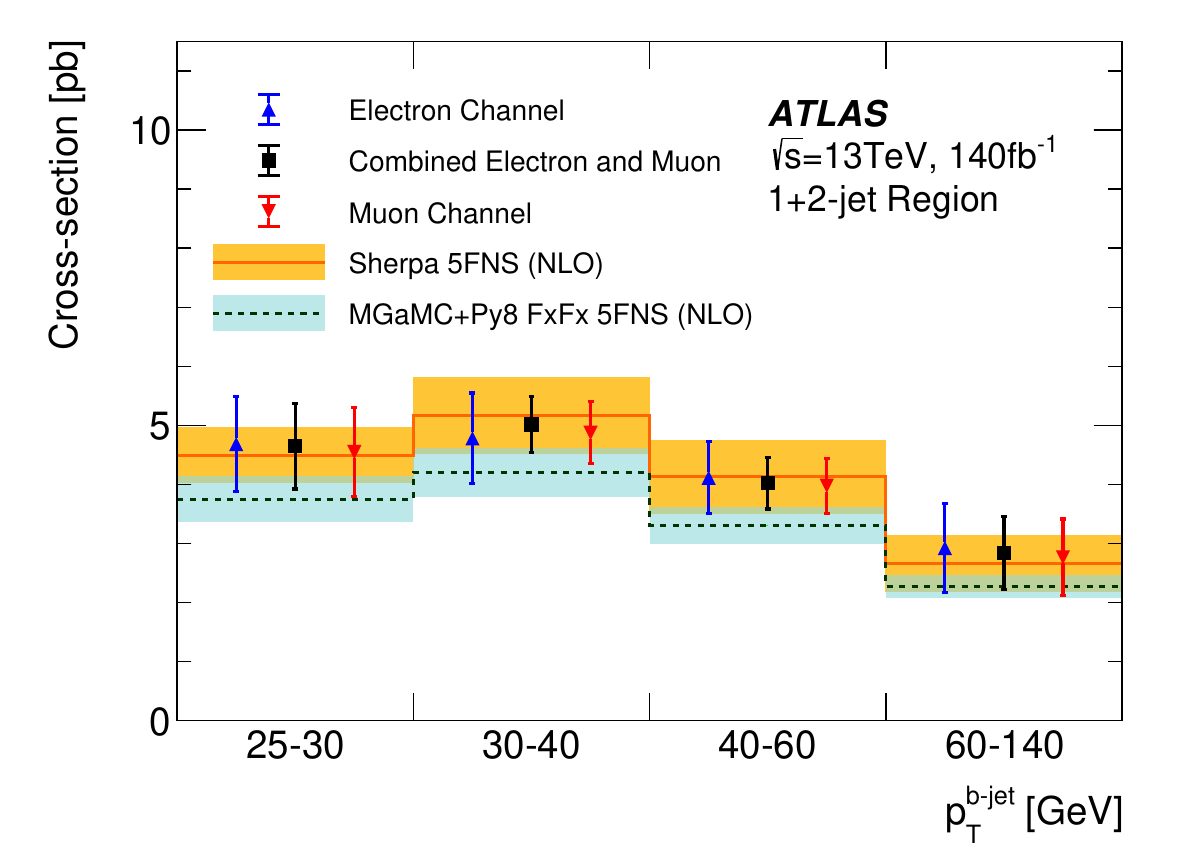}
\label{fig:BJetPtXsec_comb}
}
\\
\subfloat[]{
\includegraphics[width=0.48\textwidth]{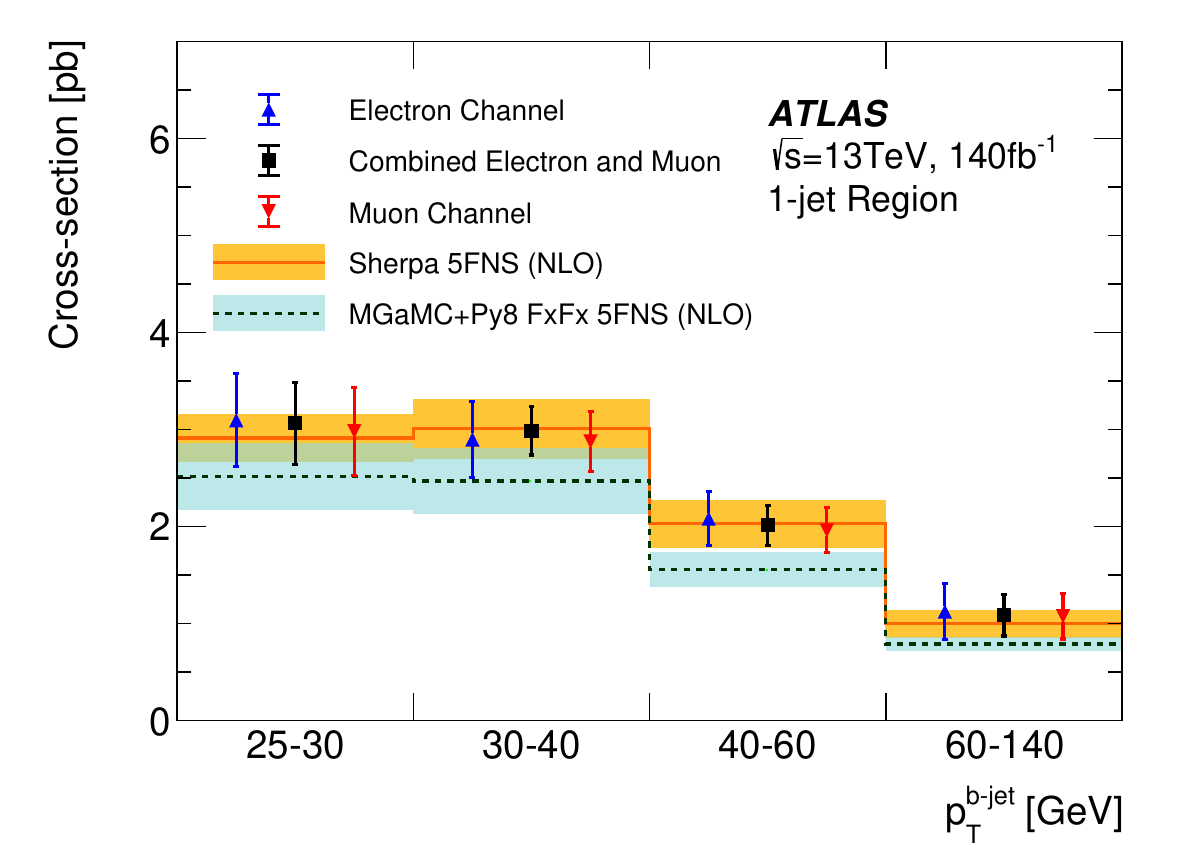}
\label{fig:BJetPtXsec_1j1t}
}
\subfloat[]{
\includegraphics[width=0.48\textwidth]{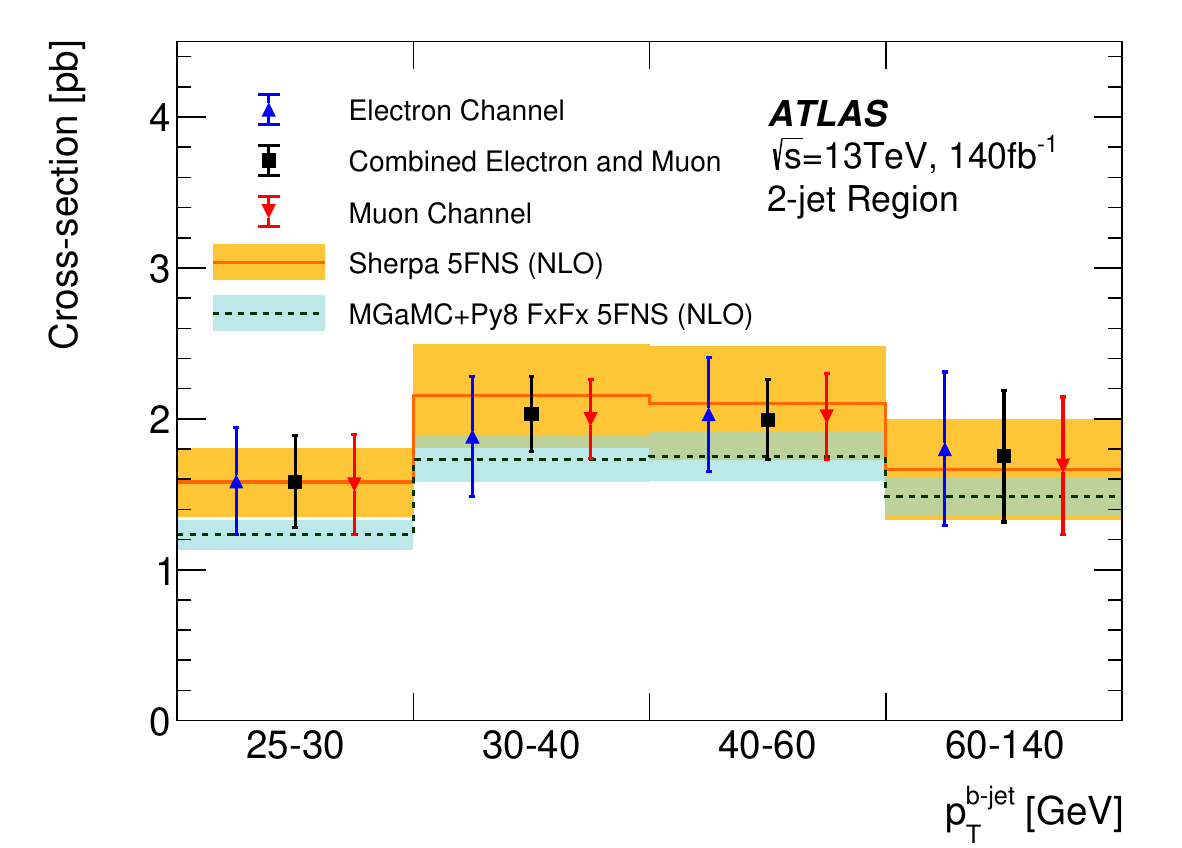}
\label{fig:BJetPtXsec_2j1t}
}

\caption{Measured cross-sections and comparisons to theoretical calculations: \protect\subref{fig:TotalXsec} total fiducial cross-sections, \protect\subref{fig:BJetPtXsec_comb} differential cross-sections versus $p_\text{T}^{\text{b-jet}}$ in the combined one and two jet region,  \protect\subref{fig:BJetPtXsec_1j1t} differential cross-sections versus $p_\text{T}^\text{b-jet}$ in the one jet region, and \protect\subref{fig:BJetPtXsec_2j1t} differential cross-sections versus $p_\text{T}^{\text{b-jet}}$ in the two jet region. The error bars of data points include both statistical and systematic uncertainties, and the error bands of theoretical predictions are from scale and PDF variations. }

\label{fig:Xsec_TheoryComparison}
\end{figure}

The unfolded results for the inclusive $W+b$-jets cross-section and the differential cross-section $\mathrm{d}\sigma/\mathrm{d}p^{\text{b-jet}}_{\text{T}}$ are shown in \cref{fig:Xsec_TheoryComparison}. The results are compared with the predictions of two event generators, \Sherpa\ and \MGNLO (see \cref{sec:dataMC}). The uncertainties on the theory predictions are from scale and PDF variations, as described in \cref{sec:uncertainties}. The dominant uncertainties are due to scale variations, which are about 10\% for both generators. The PDF uncertainties are smaller, at the level of 1--2\%. Statistical uncertainties from the finite size of the simulated samples are negligible on the theory predictions.

For the final results, the total fiducial cross-sections and the differential cross-sections in each analysis region are combined across the electron and muon channels using the Best Linear Unbiased Estimate (BLUE) method~\cite{Lyons:1988rp}. The $\chi^2 / N_{\text{dof}}$ summed over \onejoneb and \twojoneb regions, or the equivalent log likelihood $-2\log\mathcal{L}$ following the definition in Ref.~\cite{Pinto:2023yob}, are $0.2/2$ and $2.0/8$ for total cross-sections and differential cross-sections respectively, indicating good consistency between the electron and muon channel results. The statistical uncertainties are treated as uncorrelated between the two channels, while all systematic uncertainties are treated as fully correlated, except for the uncertainties associated with lepton reconstruction, identification, and trigger efficiencies, which are treated as uncorrelated.

\begin{figure}[htbp]
\centering
\includegraphics[width=0.48\textwidth]{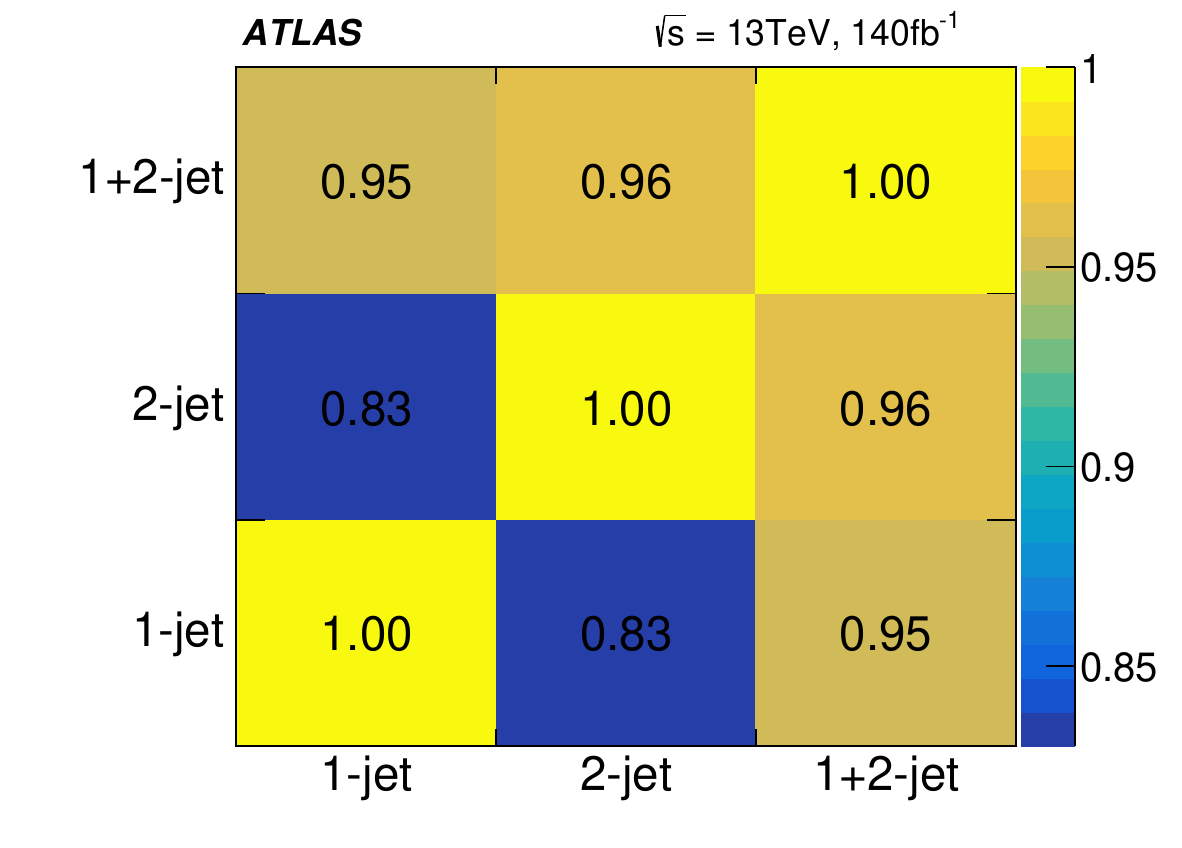}
\caption{Correlation matrix of the total uncertainties on the total fiducial cross-section.}
\label{fig:Wlnu_Corr}
\end{figure}

\begin{figure}[htbp]
\centering
\subfloat[]{
\includegraphics[width=0.48\textwidth]{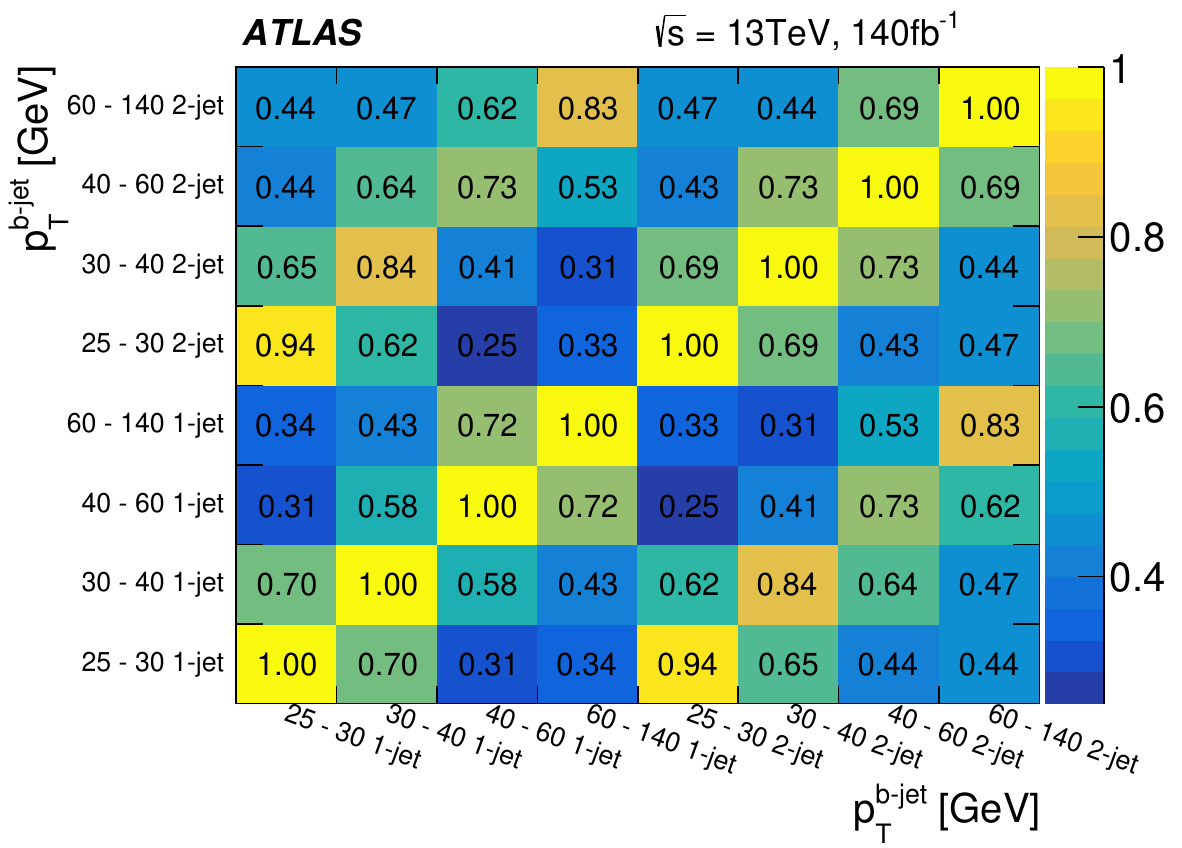}
\label{fig:Correlation_BlueCombine_1j2j}
}
\centering
\subfloat[]{
\includegraphics[width=0.48\textwidth]{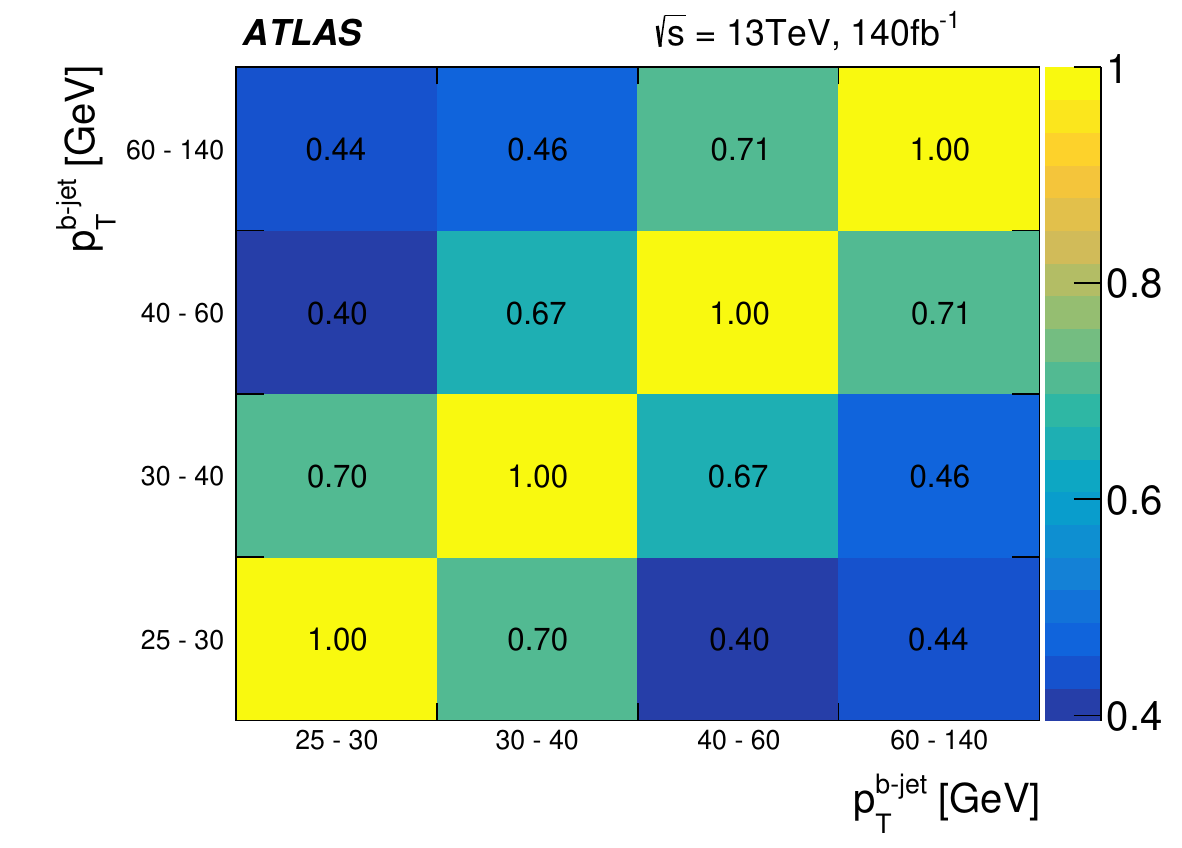}
\label{fig:Correlation_BlueCombine_comb}
}

\caption{
Correlation matrix of the total uncertainties on the combined differential cross-sections in the \protect\subref{fig:Correlation_BlueCombine_1j2j} 1-jet and 2-jet regions where the first four bins are for the 1-jet region and last four bins are for the 2-jet region, and \protect\subref{fig:Correlation_BlueCombine_comb} 1+2-jet region.
}

\label{fig:Correlation_BlueCombine}
\end{figure}

The numerical values corresponding to the combination of the electron and muon channels are shown in \cref{tab:Unfolded_Xsec_BlueCombine,tab:Unfolded_Xsec_BJetPt_1j1t_BlueCombine,tab:Unfolded_Xsec_BJetPt_2j1t_BlueCombine,tab:Unfolded_Xsec_BJetPt_comb_BlueCombine}, together with a breakdown of the systematic uncertainties.
The total fiducial cross-section measurement has an uncertainty of approximately 10\%, dominated by systematic uncertainties, while the statistical uncertainty is about 1\%.
The systematic uncertainties mainly arise from $b$-jet performance, with additional sizeable $t\bar{t}$-related contributions in the \twojoneb region due to
its higher background fraction. In all regions, the experimental measurements
show good agreement with the \Sherpa\ prediction, while the \MGNLO\ results are systematically lower by about one standard deviation. Similarly, for the
differential cross-sections, the measurements are dominated by the $b$-jet related systematic
uncertainties. However, as the $b$-jet $p_{\text{T}}$ increases, due to the kinematic distribution of the $t\bar{t}$ background, the associated
uncertainties increase significantly and gradually become dominant. The measurements in different $b$-jet $p_{\text{T}}$ bins also show
good agreement with the \Sherpa\ prediction, while the \MGNLO\ results again exhibit a systematic deficit of about one standard deviation.
The correlation matrices for the inclusive and differential measurements are presented in~\cref{fig:Wlnu_Corr} and~\cref{fig:Correlation_BlueCombine}. As this analysis is dominated by systematic uncertainties and most systematic sources are highly correlated across bins, the correlations are large.

As discussed above, these results correspond to fiducial cross-sections for $W$ boson production in association
with one or two $b$-jets. The cross-sections for $W$ boson production in association with a single $b$-jet and a
non-$b$-jet can be inferred using the following \Sherpa\ predictions for the fraction of events with one $b$-jet in the
\twojoneb fiducial region: $\{0.98, 0.92, 0.87, 0.84\}$ for the $b$-jet \pt\ ranges 25--30 GeV, 30--40 GeV, 40--60 GeV, and 60--140 GeV, respectively.

\FloatBarrier

\begin{table}[htbp]
\centering
\caption{Combined $W+b$-jet fiducial cross-sections with a breakdown of uncertainties. The unfolding uncertainty includes both the statistical uncertainty in the unfolding procedure and the modelling uncertainty. Predictions from both \textsc{Sherpa} and \textsc{MGaMC+Py8 FxFx} are shown for comparison. The uncertainties on the predictions are from scale and PDF variations.}
\begin{tabular}{l|c c c}
Region                                  & 1-jet         & 2-jets        & 1+2-jets       \\
\hline
\textsc{Sherpa} prediction~[pb]         & $9.0 \pm 0.9$ & $7.8 \pm 1.3$ & $16.8 \pm 2.3$ \\
\hline
\textsc{MGaMC+Py8 FxFx} prediction~[pb] & $7.4 \pm 0.9$ & $6.5 \pm 0.6$ & $13.9 \pm 1.3$ \\
\hline
Measured value~[pb]                     & $9.1 \pm 0.9$ & $7.5 \pm 1.1$ & $16.6 \pm 1.9$ \\
\hline
Stat. Unc. [\%]                         & 0.8           & 1.2           & 0.9            \\
Syst. Unc. [\%]                         & 10.1          & 14.7          & 11.6           \\
\hline
\multicolumn{4}{c}{Breakdown of systematic uncertainty [\%]}                             \\
\hline
Stat. Unc.~(MC)                         & 0.9           & 1.2           & 0.9            \\
$V$+jet Scale                           & 0.6           & 1.5           & 0.8            \\
$V$+jet PDF                             & 0.6           & 0.5           & 0.4            \\
Top background                          & 3.1           & 7.0           & 4.3            \\
Electron                                & 1.2           & 1.7           & 1.4            \\
Muon                                    & 2.2           & 3.3           & 2.7            \\
JVT                                     & <0.1          & <0.1          & <0.1           \\
Jet $\pt$                               & 3.5           & 3.0           & 3.1            \\
$b$-tag                                 & 7.3           & 10.2          & 8.3            \\
$\met$                                  & 2.6           & 4.7           & 3.5            \\
Luminosity                              & 1.3           & 2.2           & 1.7            \\
Multijet background                     & 3.3           & 3.2           & 3.2            \\
Unfolding                               & 0.4           & 0.6           & <0.1           \\
\hline
\end{tabular}
\label{tab:Unfolded_Xsec_BlueCombine}
\end{table}

\begin{table}[htbp]
\centering
\caption{Combined differential cross-sections in the 1-jet region with a breakdown of uncertainties. The unfolding uncertainty includes the statistical uncertainty in the unfolding procedure, the modelling uncertainty and the unfolding bias introduced by the Bayesian unfolding method. Predictions from both \textsc{Sherpa} and \textsc{MGaMC+Py8 FxFx} are shown for comparison. The uncertainties on the predictions are from scale and PDF variations.}
\begin{tabular}{l|c c c c}
$p_\text{T}^{\text{b-jet}}$~[GeV]       & [25, 30]        & [30, 40]        & [40, 60]        & [60, 140]       \\
\hline
\textsc{Sherpa} prediction~[pb]         & $2.91 \pm 0.24$ & $3.01 \pm 0.30$ & $2.03 \pm 0.25$ & $1.00 \pm 0.14$ \\
\hline
\textsc{MGaMC+Py8 FxFx} prediction~[pb] & $2.52 \pm 0.35$ & $2.47 \pm 0.34$ & $1.56 \pm 0.18$ & $0.79 \pm 0.07$ \\
\hline
Measured value~[pb]                     & $3.06 \pm 0.43$ & $2.99 \pm 0.25$ & $2.01 \pm 0.21$ & $1.09 \pm 0.21$ \\
\hline
Stat. Unc. [\%]                         & 1.6             & 1.6             & 1.8             & 2.9             \\
Syst. Unc. [\%]                         & 13.9            & 8.4             & 10.2            & 19.7            \\
\hline
\multicolumn{5}{c}{Breakdown of systematic uncertainty [\%]}                                                    \\
\hline
Stat. Unc.~(MC)                         & 2.6             & 2.0             & 1.8             & 2.0             \\
$V$+jet scale                           & 0.4             & 0.3             & 0.7             & 0.8             \\
$V$+jet PDF                             & 1.4             & 0.8             & 1.1             & 1.3             \\
Top background                          & 0.8             & 1.9             & 4.8             & 12.9            \\
Electron                                & 1.1             & 0.6             & 1.0             & 3.8             \\
Muon                                    & 1.9             & 1.6             & 2.3             & 4.5             \\
JVT                                     & <0.1            & <0.1            & <0.1            & <0.1            \\
Jet $\pt$                               & 2.8             & 3.0             & 4.4             & 7.1             \\
$b$-tag                                 & 12.4            & 6.3             & 5.2             & 6.4             \\
$\met$                                  & 3.1             & 2.3             & 3.9             & 7.7             \\
Luminosity                              & 0.9             & 1.1             & 1.7             & 3.1             \\
Multijet background                     & 1.7             & 2.1             & 2.7             & 5.2             \\
Unfolding                               & 2.4             & 0.5             & 1.1             & 1.0             \\
\hline
\end{tabular}
\label{tab:Unfolded_Xsec_BJetPt_1j1t_BlueCombine}
\end{table}

\begin{table}[htbp]
\centering
\caption{Combined differential cross-sections in the 2-jet region with a breakdown of uncertainties. The unfolding uncertainty includes the statistical uncertainty in the unfolding procedure, the modelling uncertainty and the unfolding bias introduced by the Bayesian unfolding method. Predictions from both \textsc{Sherpa} and \textsc{MGaMC+Py8 FxFx} are shown for comparison. The uncertainties on the predictions are from scale and PDF variations.}
\begin{tabular}{l|c c c c}
$p_\text{T}^{\text{b-jet}}$~[GeV]       & [25, 30]        & [30, 40]        & [40, 60]        & [60, 140]       \\
\hline
\textsc{Sherpa} prediction~[pb]         & $1.58 \pm 0.23$ & $2.15 \pm 0.34$ & $2.10 \pm 0.37$ & $1.67 \pm 0.33$ \\
\hline
\textsc{MGaMC+Py8 FxFx} prediction~[pb] & $1.23 \pm 0.10$ & $1.73 \pm 0.15$ & $1.75 \pm 0.16$ & $1.49 \pm 0.13$ \\
\hline
Measured value~[pb]                     & $1.58 \pm 0.31$ & $2.03 \pm 0.25$ & $1.99 \pm 0.27$ & $1.75 \pm 0.44$ \\
\hline
Stat. Unc. [\%]                         & 2.2             & 1.4             & 1.4             & 2.5             \\
Syst. Unc. [\%]                         & 19.3            & 12.2            & 13.3            & 24.8            \\
\hline
\multicolumn{5}{c}{Breakdown of systematic uncertainty [\%]}                                                    \\
\hline
Stat. Unc.~(MC)                         & 3.2             & 1.7             & 1.4             & 2.3             \\
$V$+jet scale                           & <0.1            & 0.3             & 0.5             & 0.5             \\
$V$+jet PDF                             & 0.9             & 0.7             & 0.6             & 1.3             \\
Top background                          & 1.6             & 4.8             & 7.7             & 16.0            \\
Electron                                & 1.4             & 0.9             & 0.7             & 2.0             \\
Muon                                    & 2.4             & 1.9             & 2.7             & 5.0             \\
JVT                                     & <0.1            & <0.1            & <0.1            & <0.1            \\
Jet $\pt$                               & 4.2             & 3.2             & 3.0             & 7.5             \\
$b$-tag                                 & 17.7            & 9.7             & 8.2             & 11.9            \\
$\met$                                  & 2.9             & 2.6             & 4.8             & 9.7             \\
Luminosity                              & 1.1             & 1.4             & 2.1             & 3.6             \\
Multijet background                     & 1.4             & 2.0             & 3.0             & 5.3             \\
Unfolding                               & 2.9             & 1.0             & 0.2             & 0.8             \\
\hline
\end{tabular}
\label{tab:Unfolded_Xsec_BJetPt_2j1t_BlueCombine}
\end{table}

\begin{table}[htbp]
\centering
\caption{Combined differential cross-sections in the 1+2-jet region with a breakdown of uncertainties. The unfolding uncertainty includes the statistical uncertainty in the unfolding procedure, the modelling uncertainty and the unfolding bias introduced by the Bayesian unfolding method. Predictions from both \textsc{Sherpa} and \textsc{MGaMC+Py8 FxFx} are shown for comparison. The uncertainties on the predictions are from scale and PDF variations.}
\begin{tabular}{l|c c c c}
$p_\text{T}^{\text{b-jet}}$~[GeV]       & [25, 30]        & [30, 40]        & [40, 60]        & [60, 140]       \\
\hline
\textsc{Sherpa} prediction~[pb]         & $4.49 \pm 0.46$ & $5.16 \pm 0.64$ & $4.13 \pm 0.62$ & $2.67 \pm 0.47$ \\
\hline
\textsc{MGaMC+Py8 FxFx} prediction~[pb] & $3.75 \pm 0.39$ & $4.20 \pm 0.42$ & $3.31 \pm 0.32$ & $2.27 \pm 0.20$ \\
\hline
Measured value~[pb]                     & $4.64 \pm 0.72$ & $5.02 \pm 0.48$ & $4.02 \pm 0.44$ & $2.84 \pm 0.62$ \\
\hline
Stat. Unc. [\%]                         & 1.6             & 1.4             & 1.3             & 1.9             \\
Syst. Unc. [\%]                         & 15.6            & 9.5             & 10.9            & 21.7            \\
\hline
\multicolumn{5}{c}{Breakdown of systematic uncertainty [\%]}                                                    \\
\hline
Stat. Unc.~(MC)                         & 2.6             & 1.7             & 1.3             & 1.8             \\
$V$+jet Scale                           & 0.4             & 0.4             & 0.5             & 0.6             \\
$V$+jet PDF                             & 1.3             & 0.7             & 0.4             & 1.1             \\
Top background                          & 0.9             & 3.0             & 5.8             & 13.8            \\
Electron                                & 1.2             & 0.5             & 0.7             & 2.1             \\
Muon                                    & 2.1             & 1.7             & 2.4             & 4.7             \\
JVT                                     & <0.1            & <0.1            & <0.1            & <0.1            \\
Jet $\pt$                               & 2.9             & 2.7             & 3.3             & 7.0             \\
$b$-tag                                 & 14.2            & 7.6             & 6.2             & 9.6             \\
$\met$                                  & 3.0             & 2.3             & 4.4             & 8.8             \\
Luminosity                              & 0.9             & 1.2             & 1.9             & 3.4             \\
Multijet background                     & 1.6             & 2.0             & 2.7             & 5.3             \\
Unfolding                               & 2.6             & 0.7             & 0.6             & 0.9             \\
\hline
\end{tabular}
\label{tab:Unfolded_Xsec_BJetPt_comb_BlueCombine}
\end{table}

\FloatBarrier


\section{Conclusions}
\label{sec:conclusion}
A measurement of the cross-section of $W$ boson production in association with $b$-jets at $\sqrt{s}=13$ TeV is presented, based on data corresponding to an integrated luminosity of 140~$\text{fb}^{-1}$ collected with the ATLAS detector at the LHC in Run 2. The analysis defines two signal regions with exactly one or two jets in the final state, requiring exactly one of the jets to be $b$-tagged. The signal yields are estimated separately in the electron and muon decay channels and unfolded to a common fiducial phase space. Good agreement is observed between the results in the electron and muon channels, and combined measurements are provided.

The measured fiducial cross-sections in the 1-jet, 2-jets and 1+2 jets combined regions are:
\begin{align*}
& \sigma_\text{fid}(\text{1-jet}) =     9.1 \pm 0.1 \text{ (stat.) } \pm 0.9 \text{ (syst.) pb},  \\
& \sigma_\text{fid}(\text{2-jets}) =    7.5 \pm 0.1 \text{ (stat.) } \pm 1.1 \text{ (syst.) pb},  \\
& \sigma_\text{fid}(\text{1+2 jets}) =  16.6 \pm 0.2 \text{ (stat.) } \pm 1.9 \text{ (syst.) pb}.
\end{align*}

They are found to be consistent with the \Sherpa\ predictions, while while being larger than the \MGNLO\ predictions by approximately one standard deviation. The uncertainty on the total cross-section is about 10\% and is dominated by systematic uncertainties. The precision of theory predictions is limited by missing higher-order uncertainties, which are estimated through scale variations to be approximately at the 10\% level.

A differential cross-section measurement as a function of the leading $b$-jet $p_\text{T}$ is also presented, for jets with $p_\text{T}$ in the range of 25 GeV to 140 GeV. The measured differential cross-sections show good agreement with the theoretical predictions.

Compared to the previous ATLAS 7 TeV measurement~\cite{STDM-2012-11}, both statistical and systematic uncertainties are significantly reduced. These improvements are primarily driven by 30 times more data and improved jet and $b$-tagging performance.


\section*{Acknowledgements}

%
%

%
%

We thank CERN for the very successful operation of the LHC and its injectors, as well as the support staff at
CERN and at our institutions worldwide without whom ATLAS could not be operated efficiently.

The crucial computing support from all WLCG partners is acknowledged gratefully, in particular from CERN, the ATLAS Tier-1 facilities at TRIUMF/SFU (Canada), NDGF (Denmark, Norway, Sweden), CC-IN2P3 (France), KIT/GridKA (Germany), INFN-CNAF (Italy), NL-T1 (Netherlands), PIC (Spain), RAL (UK) and BNL (USA), the Tier-2 facilities worldwide and large non-WLCG resource providers. Major contributors of computing resources are listed in Ref.~\cite{ATL-SOFT-PUB-2026-001}.

We gratefully acknowledge the support of ANPCyT, Argentina; YerPhI, Armenia; ARC, Australia; BMWFW and FWF, Austria; ANAS, Azerbaijan; CNPq and FAPESP, Brazil; NSERC, NRC and CFI, Canada; CERN; ANID, Chile; CAS, MOST and NSFC, China; Minciencias, Colombia; MEYS CR, Czech Republic; DNRF and DNSRC, Denmark; IN2P3-CNRS and CEA-DRF/IRFU, France; SRNSFG, Georgia; BMFTR, HGF and MPG, Germany; GSRI, Greece; RGC and Hong Kong SAR, China; ICHEP and Academy of Sciences and Humanities, Israel; INFN, Italy; MEXT and JSPS, Japan; CNRST, Morocco; NWO, Netherlands; RCN, Norway; MNiSW, Poland; FCT, Portugal; MNE/IFA, Romania; MSTDI, Serbia; MSSR, Slovakia; ARIS and MVZI, Slovenia; DSI/NRF, South Africa; MICIU/AEI, Spain; SRC and Wallenberg Foundation, Sweden; SERI, SNSF and Cantons of Bern and Geneva, Switzerland; NSTC, Taipei; TENMAK, T\"urkiye; STFC/UKRI, United Kingdom; DOE and NSF, United States of America.

Individual groups and members have received support from BCKDF, CANARIE, CRC and DRAC, Canada; CERN-CZ, FORTE and PRIMUS, Czech Republic; COST, ERC, ERDF, Horizon 2020 and Marie Sk{\l}odowska-Curie Actions, European Union; Investissements d'Avenir Labex, Investissements d'Avenir Idex and ANR, France; DFG and AvH Foundation, Germany; Herakleitos, Thales and Aristeia programmes co-financed by EU-ESF and the Greek NSRF, Greece; BSF-NSF and MINERVA, Israel; NCN and NAWA, Poland; La Caixa Banking Foundation, CERCA and AGAUR programs from Generalitat de Catalunya and PROMETEO and GenT Programmes Generalitat Valenciana, Spain; G\"{o}ran Gustafssons Stiftelse, Sweden; The Royal Society and Leverhulme Trust, United Kingdom; Eric and Wendy Schmidt Fund for Strategic Innovation, United States of America.

In addition, individual members wish to acknowledge support from Chile: Agencia Nacional de Investigaci\'on y Desarrollo (ANID FONDECYT reg. 1230987, FONDECYT 1230812, FONDECYT 1240864, Fondecyt 3240661, Fondecyt Regular 1240721); China: Chinese Ministry of Science and Technology (MOST-2023YFA1605700, MOST-2023YFA1609300), National Natural Science Foundation of China (NSFC 12275265, NSFC-W2543005); Czech Republic: Czech Science Foundation (GACR - 24-11373S), Ministry of Education Youth and Sports (ERC-CZ-LL2327, FORTE CZ.02.01.01/00/22\_008/0004632); EU: H2020 European Research Council (ERC - 101002463); European Union: European Research Council (BARD No. 101116429, ERC - 101219398, ERC 101089007), European Regional Development Fund (HE COFUND GA No.101081355, ERDF), Marie Sklodowska-Curie Actions (GAP-101168829); France: Agence Nationale de la Recherche (ANR-21-CE31-0013, ANR-22-EDIR-0002, ANR-24-CE31-0504-01); Germany: Deutsche Forschungsgemeinschaft (DFG - 469666862); China: Research Grants Council (GRF); Italy: Istituto Nazionale di Fisica Nucleare (LHC-MIUR - 28003/2025), Ministero dell'Università e della Ricerca (NextGenEU I53D23000820006 M4C2.1.1, SOE2024\_0000023); Japan: Japan Society for the Promotion of Science (JSPS KAKENHI  JP25H0063, JSPS KAKENHI JP22H04944, JSPS KAKENHI JP24K23939, JSPS KAKENHI JP24KK0251, JSPS KAKENHI JP25H00650, JSPS KAKENHI JP25H01291, JSPS KAKENHI JP25K01011, JSPS KAKENHI JP25K01023, JSPS KAKENHI JP25KK0047); Poland: Polish National Science Centre (NCN 2021/42/E/ST2/00350, NCN OPUS 2023/51/B/ST2/02507, NCN OPUS nr 2022/47/B/ST2/03059, NCN UMO-2019/34/E/ST2/00393, UMO-2022/47/O/ST2/00148, UMO-2023/49/B/ST2/04085, UMO-2023/51/B/ST2/00920, UMO-2024/53/N/ST2/00869); Spain: Agència de Gestió d'Ajuts Universitaris i de Recerca. (AGAUR - 2023 BP 00141), Ministry of Science and Innovation (RYC2019-028510-I, RYC2020-030254-I, RYC2021-031273-I, RYC2022-038164-I), Ministerio de Ciencia, Innovación y Universidades/Agencia Estatal de Investigaci\'on (EU NextGenerationEU (PRTR-C17.I1), PID2022-142604OB-C22); Sweden: Carl Trygger Foundation (Carl Trygger Foundation CTS 22:2312), Swedish Research Council (Swedish Research Council 2023-04654, VR 2021-03651, VR 2022-03845, VR 2022-04683, VR 2023-03403, VR 2024-05451, VR 2025-05940), Knut and Alice Wallenberg Foundation (KAW 2023.0366); United Kingdom: The Binks Trust, Royal Society (NIF-R1-231091); United States of America: U.S. Department of Energy (ECA DE-AC02-76SF00515), John Templeton Foundation (John Templeton Foundation 63206), Neubauer Family Foundation.

%
%


\clearpage
\printbibliography

\clearpage
 
\begin{flushleft}
\hypersetup{urlcolor=black}
{\Large The ATLAS Collaboration}

\bigskip

\AtlasOrcid[0000-0002-6665-4934]{G.~Aad}$^\textrm{\scriptsize 102}$,
\AtlasOrcid[0000-0001-7616-1554]{E.~Aakvaag}$^\textrm{\scriptsize 17}$,
\AtlasOrcid[0000-0002-5888-2734]{B.~Abbott}$^\textrm{\scriptsize 121}$,
\AtlasOrcid[0000-0002-0287-5869]{S.~Abdelhameed}$^\textrm{\scriptsize 83b}$,
\AtlasOrcid[0000-0002-1002-1652]{K.~Abeling}$^\textrm{\scriptsize 54}$,
\AtlasOrcid[0000-0001-5763-2760]{N.J.~Abicht}$^\textrm{\scriptsize 48}$,
\AtlasOrcid[0000-0002-8496-9294]{S.H.~Abidi}$^\textrm{\scriptsize 30}$,
\AtlasOrcid[0009-0003-6578-220X]{M.~Aboelela}$^\textrm{\scriptsize 44}$,
\AtlasOrcid[0000-0002-9987-2292]{A.~Aboulhorma}$^\textrm{\scriptsize 36e}$,
\AtlasOrcid[0000-0001-5329-6640]{H.~Abramowicz}$^\textrm{\scriptsize 154}$
% ...
\bigskip
\\

% ...
$^{17}$Department for Physics and Technology, University of Bergen, Bergen; Norway.\\
% ...
$^{30}$Physics Department, Brookhaven National Laboratory, Upton NY; United States of America.\\
% ...
$^{36}$$^{(a)}$Facult\'e des Sciences Ain Chock, Universit\'e Hassan II de Casablanca;$^{(b)}$Facult\'{e} des Sciences, Universit\'{e} Ibn-Tofail, K\'{e}nitra;$^{(c)}$Facult\'e des Sciences Semlalia, Universit\'e Cadi Ayyad, LPHEA-Marrakech;$^{(d)}$LPMR, Facult\'e des Sciences, Universit\'e Mohamed Premier, Oujda;$^{(e)}$Facult\'e des sciences, Universit\'e Mohammed V, Rabat;$^{(f)}$Institute of Applied Physics, Mohammed VI Polytechnic University, Ben Guerir; Morocco.\\
% ...
$^{44}$Physics Department, Southern Methodist University, Dallas TX; United States of America.\\
% ...
$^{48}$Fakult\"{a}t Physik, Technische Universit{\"a}t Dortmund, Dortmund; Germany.\\
% ...
$^{54}$II. Physikalisches Institut, Georg-August-Universit\"{a}t G\"ottingen, G\"ottingen; Germany.\\
% ...
$^{83}$$^{(a)}$Khalifa University of Science and Technology, Abu Dhabi;$^{(b)}$New York University Abu Dhabi, Abu Dhabi;$^{(c)}$United Arab Emirates University, Al Ain;$^{(d)}$University of Sharjah, Sharjah; United Arab Emirates.\\
% ...
$^{102}$CPPM, Aix-Marseille Universit\'e, CNRS/IN2P3, Marseille; France.\\
% ...
$^{121}$Homer L. Dodge Department of Physics and Astronomy, University of Oklahoma, Norman OK; United States of America.\\
% ...
$^{154}$Raymond and Beverly Sackler School of Physics and Astronomy, Tel Aviv University, Tel Aviv; Israel.\\
% ...
$^{a}$ Also at Affiliated with an institute formerly covered by a cooperation agreement with CERN.\\
$^{b}$ Also at An-Najah National University, Nablus; Palestine.\\
$^{c}$ Also at Borough of Manhattan Community College, City University of New York, New York NY; United States of America.\\
$^{d}$ Also at Center for Interdisciplinary Research and Innovation (CIRI-AUTH), Thessaloniki; Greece.\\
$^{e}$ Also at Centre of Physics of the Universities of Minho and Porto (CF-UM-UP); Portugal.\\
$^{f}$ Also at CERN, Geneva; Switzerland.\\
% ...

\end{flushleft}


\clearpage

\end{document}